\documentclass[11pt,a4paper]{article}

\usepackage{amsmath, amsthm, amssymb, mathrsfs, bm}    
\usepackage{color}      
\usepackage{graphicx}   
\usepackage{subcaption} 
\usepackage{float} 
\usepackage[square,sort,comma,numbers]{natbib}    
\usepackage{palatino}   
\usepackage{pgfplots} \pgfplotsset{compat=newest}
\usetikzlibrary{calc}
\usepackage{tkz-euclide}
\usepackage[left=2cm,right=2cm,top=2.75cm,bottom=2.75cm]{geometry}
\usepackage[compact]{titlesec}
\usepackage{fancyhdr} 
\usepackage{abstract}

\usepackage[]{authblk}

\newcommand{\infint}{\int_{-\infty}^{\infty}}
\newcommand{\pdev}[2]{ \frac{\partial {#1}}{\partial {#2}} } 
\newcommand{\inflim}{\lim_{x \rightarrow {\infty}} u(x,t) }
\newcommand{\neginflim}{\lim_{x \rightarrow {-\infty}} u(x,t) }
\newcommand{\real}{$\mathbb{R}$}
\newcommand{\field}{\(\text{\real $\times$ \real} \rightarrow \text{\real}\)}
\newcommand{\intu}{\int_{u_{0}}^{u_{1}}}
\newcommand{\lnn}[1]{\ln\left(#1\right)}

\newcommand{\sech}{{\rm sech}}

\numberwithin{equation}{section}

\title{Solitons: Kinks, Collisions and Breathers}
\author{Shivani Sickotra}
\affil{University of Sheffield, University of Leeds}
\date{March 20, 2020}

\begin{document}
\maketitle

\begin{abstract}
In this report, the various 1D single soliton and multi-soliton solutions of the Sine-Gordon equation are explored. First the topological kink solitons and their properties for the Sine-Gordon, as well as the $\phi^{4}$ model are investigated. This includes energy and momentum conservation laws for both models, as well as static kinks and antikinks. The Bogomol'nyi equation and the importance of Lorentz invariance are also discussed. Then in the second half, the focus is on Sine-Gordon multi-kink solutions, which are found by a separation of variables method rather than using inverse scattering or B\"acklund transforms. These kink-kink and kink-antikink collisions are explained by considering topological charges and interaction energies. The special bound state dynamical breather soliton is obtained and properties such as the binding energy of the breather are presented. Following this, the Frenkel-Kontorova model is explored, leading to the discrete Sine-Gordon model and the introduction of discrete breathers.
\end{abstract}

\setcounter{page}{1}
\tableofcontents
\medskip
\section{Introduction}
\label{s:intro} 
Solitons are a phenomenon that arise in nonlinear waves and they can be described by the solutions of nonlinear partial differential equations (PDE's). They are smooth, localised solutions and so they can be thought of as solitary waves. There is a mathematical difference between the term solitary waves and solitons. The former refers to localised solutions of non-integrable equations, whereas the latter refers to localised solutions of integrable equations \citep{yousefi2012}. The models that we focus on are the Sine-Gordon(SG) and $\phi^{4}$ models which are both integrable, and so we refer to the solutions as ‘solitons’.

The first observation of a soliton was by J. Scott Russell in 1834 \citep{Russell1845}, when he witnessed a travelling vessel in a canal that suddenly stopped. Because of this motion, the waves in the canal accumulated around the vessel bow and he saw that they propelled forward as one individual wave. He noticed this solitary wave could travel large distances whilst maintaining a constant speed and amplitude, naming this phenomenon as the ‘Wave of translation’. 

In particular, an interesting property of solitons is that they behave in an analogous way to particles and hence support wave-particle duality, which is reminiscent of quantum mechanics. They have a mass, conserve energy and momentum as well as interact with each other in the same way that elementary particles of matter and antimatter do. Solitons exist because of a balance between nonlinearity and dispersion\citep{yousefi2012} and because of this they are stable, robust and retain their profile even during collisions, much like two particles would. The integrability of the SG model allows these conservation laws to hold and it turns out that the SG model is also Lorentz invariant. The ending ‘-on’ in the name soliton stems from the idea that the wave solutions of integrable equations behave as relativistic particles. Since Russel’s first scientific observation, solitons have been recognised in nerve pulses, DNA, tornados, fibre optics and many other physical applications \citep{rem1999,daux2006}.
\medskip \\

\section{Klein-Gordon Equation}
\label{s:kgequ}
Solitons can be classified in numerous ways, such as considering their topology or their profiles. The solitons that will be focused on are called the Kink and Breather soliton. The kink soliton is topological, meaning that the boundary conditions at infinity for the wave are topologically different to the vacuum it is in \citep{yousefi2012}. These kink solitons are also characterised by their permanent profiles, which mean that they do not change with time. The breather profile however is internally dynamic and can depend on time. These solitons can be described by the solution to the {\bf nonlinear Klein-Gordon} PDE
\begin{equation} 
u_{tt}-u_{xx}+V'(u) = 0, 
\label{kgequ}
\end{equation} 
where $V$ represents the {\bf potential} and $V'$ is the derivative of $V$. The independent variable $x$ represents a one dimensional position in space and $t$ represents time. Any function $u$ : \field\: is called a {\bf field} \citep{fitzgerald2019}. A {\bf vacuum} of the potential function $V$, in \eqref{kgequ} is any value $u_{0}\in \mathbb{R}$ where $\bm{V(u_{0}) = 0}$. \smallskip \\

\noindent Consider the constant field $u(x,t) = u_{0}$, where $u_{0}$ is a vacuum of the potential $V$. It can be shown that this satisfies the PDE \eqref{kgequ} as the second order partial derivatives of $u(x,t) = u_{0}$ with respect to $x$ and $t$ are both zero. As $u_{0}$ is a vacuum, it then follows that $V(u_{0}) = 0$, and so $V'(u_{0}) = 0$. Substituting these into \eqref{kgequ}, we obtain $0 = 0$, hence $u(x,t) = u_{0}$ is a solution of the Klein-Gordon equation \eqref{kgequ}. These types of solutions are called {\bf vacuum solutions}.

\section{Sine-Gordon Equation}
\label{s:s:sG} 
If a smooth function $V$ : {\(\text{\real} \rightarrow \text{\real}\)}, such as $V(u) = 1 - \cos(u),$ is considered, then this particular potential function equals zero only at a discrete set of points and is also non negative, so $V(u) \geq 0 \; \text{for all} \; u \in \text{\real}$ \citep{fitzgerald2019}.\\ 
\newline If this potential function is used in the Klein-Gordon equation \eqref{kgequ}, then the resulting equation is known as the {\bf Sine-Gordon equation} (SG equation)
\begin{equation}
u_{tt}-u_{xx}+\sin(u) = 0
\label{sGequ}
\end{equation}
Using the ansatz $u(x,t) = C$ in \eqref{sGequ}, where $C$ is a constant, we obtain $\sin(C) =0$. This is solved to get the constant solution, $u_{n}(x,t) = C = 2\pi n$, where $n \in \mathbb{Z}$.
These are the trivial solutions of the SG equation \eqref{sGequ}, also known as the vacuum solutions.

\subsection{Linearisation of the Sine-Gordon Equation}
\label{s:linsg}
We can linearise \eqref{sGequ} by first supposing $u$ is very small, $|u| \ll 1$. Then we know that the Taylor series expansion of $\sin(u)$ can be written as 
\begin{equation*}
\sin(u) = u - \frac{u^{3}}{3!} + \frac{u^{5}}{5!} - ...
\end{equation*}
and so \eqref{sGequ} can be written as 
\begin{equation}
u_{tt} - u_{xx} \approx -u + \frac{u^{3}}{3!} - \frac{u^{5}}{5!} + ...
\label{taylorsineg}
\end{equation}
If we use the 1D complex travelling wave solution $u(x,t) = Ae^{i(\kappa x - \omega t)}$ as an ansatz, we can calculate the second derivatives of this with respect to $x$ and $t$ and substitute these into \eqref{taylorsineg} to obtain,
\begin{equation*}
-A\omega^{2}e^{i(\kappa x - \omega t)} + A\kappa^{2}e^{i(\kappa x - \omega t)} = -Ae^{i(\kappa x - \omega t)} + \frac{A^{3}e^{3i(\kappa x - \omega t)}}{3!} - \frac{A^{5}e^{5i(\kappa x - \omega t)}}{5!} + ...
\end{equation*}
We can linearise this, provided $u$ remains small, by discarding all higher order terms, 
\begin{equation*}
-Ae^{i(\kappa x - \omega t)} \left(\omega^{2} - \kappa^{2}\right) = -Ae^{i(\kappa x - \omega t)}
\end{equation*}
Simplifying this leaves us with,
\begin{equation}
\omega_{\kappa}^{2} = 1 + \kappa^{2}
\label{clindisprel}
\end{equation}
which is known as the dispersion relation between angular frequency $\omega_{\kappa}^{2}$ and wave number $\kappa$.  \smallskip \\
Therefore, $u(x,t) \sim e^{i(\kappa x - \omega t)}$ is a solution of the linearised SG equation. These small amplitude linear waves are called {\bf phonons}\citep{benner2000} and they are the simplest nontrivial solutions and they are one out of three elementary excitations of the SG model. The two remaining elementary excitations are kinks and breather solitons, which will be the focus of the following sections. 

\section{Kink Solitons}
\subsection{Kinks and Anitikinks}
If a potential $V$ is assumed to have more than one vacuum, and that $u_{0}, u_{1}$ with $u_{0} < u_{1}$ are neighbouring vacua, then a {\bf kink} is any solution of \eqref{kgequ} with boundary behaviour 
\begin{equation}
\neginflim = u_{0}, \hspace{10mm}
\inflim = u_{1} 
\label{kinkbound}
\end{equation}
Similarly, an {\bf antikink} is any solution of \eqref{kgequ} with boundary behaviour 
\begin{equation}
\neginflim = u_{1}, \hspace{10mm}
\inflim = u_{0} 
\label{akinkbound}
\end{equation}

\begin{figure}[H] \begin{center} 
\begin{tikzpicture}
\begin{axis}[xmin=-2.5, xmax=2.5,
    ymin=-1.5, ymax=1.5, samples=1000,
    axis lines=center,
    axis on top=true,
    domain=-2.5:2.5,
    xlabel=$x$, ylabel =$u(x)$, xtick={0}, ytick={0}, legend pos=outer north east]
\addplot+[no marks, red, thick] {tanh(x)};
\addlegendentry{Kink}
\addplot+[no marks, teal,thick] {-tanh(x)};
\addlegendentry{Antikink}
\addplot+[no marks, dotted, thick]{-1};  \node [right, red] at (axis cs: -2.5,-1.1) {$u_{0}$}; \node [right, teal] at (axis cs: -2.5,-1.2) {$u_{1}$};
\addplot+[no marks, dotted, thick]{1}; \node [right, red] at (axis cs: -2.5,1.1) {$u_{1}$}; \node [right, teal] at (axis cs: -2.5,1.2) {$u_{0}$};
\end{axis} \end{tikzpicture}
 \end{center}
\vspace{-5mm} \caption{The shape of a general kink and an antikink with boundary behaviour illustrated respectively by asymptotes $u_0$ and $u_1$.}
\label{genkink}
\end{figure}
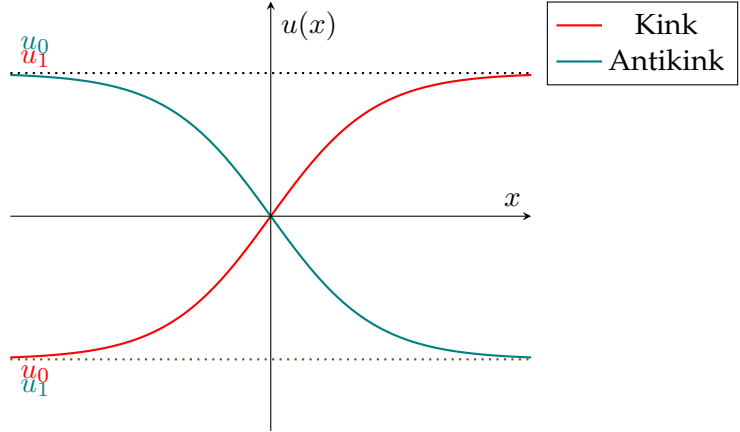

\medskip
\noindent
If $u_{K}$ : \field \: is a kink solution of \eqref{kgequ}, then it can be shown that $u_{\bar{K}}$ : \field \: defined by $$u_{\bar{K}}(x,t) = u_{K}(-x,t)$$ is an antikink solution of \eqref{kgequ}.

\begin{proof}
We have $u_{\bar{K}}(x,t) = u_{K}(-x,t)$ and as $t$ remains unchanged, then $u_{\bar{K}_{tt}}(x,t) = u_{K_{tt}}(-x,t)$.
\medskip Define  $\bm{y = - x}$ and hence note that $\frac{dy}{dx} = - 1$.\\
Then we can find $ u_{\bar{K}_{x}}$ and $u_{\bar{K}_{xx}}$ using the chain rule,
\begin{align*}
u_{\bar{K}_{x}}(x,t) &=\: \pdev{u_{\bar{K}}(x,t)}{x} =\: \pdev{u_{K}(y,t)}{y} \frac{dy}{dx} =\:  -\pdev{u_{K}(y,t)}{y}\\
u_{\bar{K}_{xx}}(x,t) &=\: \pdev{}{x}\left( -\pdev{u_{K}(y,t)}{y}\right) \frac{dy}{dx} =\: \pdev{}{x}\left(\pdev{u_{K}(y,t)}{y}\right) =\: \pdev{}{x}\left(\pdev{u_{K}(-x,t)}{x}\right) =\: u_{{K}_{xx}}(-x,t)
\end{align*}
Therefore we have, $u_{\bar{K}_{tt}}(x,t) = u_{K_{tt}}(-x,t)$ and $u_{\bar{K}_{xx}}(x,t) = u_{{K}_{xx}}(-x,t)$, so equation \eqref{kgequ} becomes $u_{K_{tt}}(-x,t) - u_{{K}_{xx}}(-x,t) + V'(u) = 0$. As $x$ is negative, the solution of the equation has boundary behaviour corresponding to \eqref{akinkbound}, hence $u_{\bar{K}}(x,t) = u_{K}(-x,t)$ is an antikink solution to \eqref{kgequ}.
\end{proof}

\subsubsection{Sine-Gordon Kinks}
\label{s:s:sgkinks}
A solution of the SG equation \eqref{sGequ} is the function,
\begin{equation}
u(x,t)  = 4\:\tan^{-1}e^{\gamma(x-vt)}
\label{sGsol}
\end{equation} 
where velocity $v \in (-1, 1)$ is a constant and
\begin{equation} 
\gamma = (1-v^{2})^{-1/2}
\label{gamma}
\end{equation}

\begin{proof}
The second order partial derivatives for \eqref{sGsol} with respect to $x$ and $t$ can be found as, 
\begin{equation*}
u_{xx} = \frac{4\gamma(\gamma e^{\gamma (x-vt)}-\gamma e^{3\gamma (x-vt)})}{(e^{2\gamma(x-vt)}+1)^{2}} \quad \quad
u_{tt} = \frac{-4\gamma v(\gamma v e^{3\gamma (x-vt)}-\gamma v e^{\gamma (x-vt)})}{(e^{2\gamma(x-vt)}+1)^{2}}
\end{equation*}
To find $V'(u) = \sin(u) = \sin(4\tan^{-1}e^{\gamma(x-vt)})$, we can first use the Pythagoras theorem. Define 
\begin{equation}
z = e^{\gamma(x-vt)} \hspace{5mm} \& \hspace{5mm} y = \tan^{-1}(z)
\label{definezy}
\end{equation}
Therefore, $z = \tan(y)$ and we can construct the right angled triangle
\begin{figure}[H]  \begin{center}
\begin{tikzpicture}[thick, scale=0.8]
\coordinate (O) at (0,0);
\coordinate (A) at (4,0);
\coordinate (B) at (0,2);
\draw (O)--(A)--(B)--cycle;
\tkzLabelSegment[below=2pt](O,A){$1$}
\tkzLabelSegment[left=2pt](O,B){$z$}
\tkzLabelSegment[above right=2pt](A,B){$\sqrt{z^{2}+1}$}
\tkzMarkRightAngle[fill=orange,size=0.3,opacity=.4](A,O,B)
\tkzMarkAngle[fill=orange,size=1cm,opacity=.4](B,A,O)
\tkzLabelAngle[pos = 1.3](B,A,O){$y$}
\end{tikzpicture} \end{center} \end{figure} \vspace{-3mm}
\noindent From this we find,
\begin{equation}
\sin(y) = \frac{z}{\sqrt{z^{2}+1}} \hspace{5mm} \& \hspace{5mm} \cos(y) = \frac{1}{\sqrt{z^{2}+1}}
\label{sincosy}
\end{equation}
Now, using the double angle formulas $\sin(2x) = 2\sin(x)\cos(x)$ and $\cos(2x) = 2\cos^{2}(x) - 1$, we can write
\begin{equation*}
\sin(4y) = 2\sin(2y)\cos(2y) = 8\sin(y)\cos^{3}(y) - 4\sin(y)\cos(y)
\end{equation*}
We can substitute \eqref{sincosy} and simplify this to obtain
\begin{equation*}
\sin(4y) = \frac{8z}{(z^{2}+1)^{2}} - \frac{4z}{z^{2}+1} = \frac{4z - 4z^{3}}{(z^{2}+1)^{2}}
\end{equation*}
Substituting back $z$, we find that,
$$\sin(u) = \frac{4(e^{\gamma (x-vt)}-e^{3\gamma (x-vt)})}{(e^{2\gamma(x-vt)}+1)^{2}}$$
Substituting the expressions for $u_{xx}$, $u_{tt}$ and $V'(u)$ into \eqref{sGequ}, we find that $0 = 0$, and hence \eqref{sGsol} is a solution of the Sine-Gordon equation.  
\end{proof} 
\noindent In fact, \eqref{sGsol} is a {\bf kink} solution with vacua $u_{0}=0$ and $u_{1}=2\pi$. As vacua occur when $V(u_{*}) = 1 - \cos(u_{*}) = 0$, this means that the values bounding the kink must be $u_{0}=0$ and $u_{1}=2\pi$ in order to satisfy this equation.
Furthermore, it can be seen that \eqref{sGsol} also exhibits the boundary behaviour \eqref{kinkbound}
\begin{align}
u_{0} &= \neginflim = \lim_{x \rightarrow {-\infty}}4\tan^{-1}e^{\gamma(x-vt)} = 4 \times 0 = 0 \nonumber \\
u_{1} &= \inflim = \lim_{x \rightarrow {\infty}}4\tan^{-1}e^{\gamma(x-vt)} = 4 \times \frac{\pi}{2} = 2\pi
\label{sgkinklim}
\end{align}
Therefore, is it a kink solution of the Sine-Gordon equation.
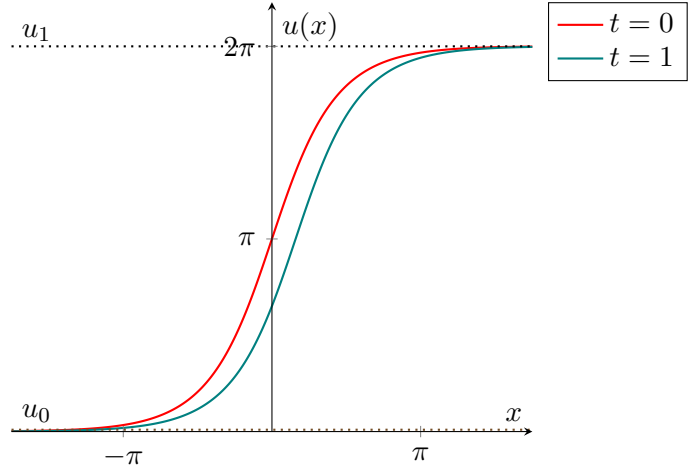
\begin{figure}[H] \begin{center}
\begin{tikzpicture}
\begin{axis}[xmin=-5.5, xmax=5.5,
    ymin=0, ymax=7, samples=1000,
    axis lines=center,
    axis on top=true,
    domain=-5.5:5.5,
    xlabel=$x$, ylabel =$u(x)$, xtick={-3.14, 0, 3.14}, ytick={3.14, 6.28},  xticklabels ={$-\pi$, 0, $\pi$}, yticklabels = {$\pi$, $2\pi$}, legend pos=outer north east]
\addplot+[no marks, red, thick] {4*rad(atan(exp(2*sqrt(3)*x/3)))};
\addlegendentry{$t=0$}
\addplot+[no marks, teal,thick] {4*rad(atan(exp(2*sqrt(3)*(x-1/2)/3)))};
\addlegendentry{$t=1$}
\addplot+[no marks, dotted, thick]{0.03};  \node [right] at (axis cs: -5.5,0.3) {$u_{0}$};
\addplot+[no marks, dotted, thick]{6.28}; \node [right] at (axis cs: -5.5,6.5) {$u_{1}$};
\end{axis}
\end{tikzpicture}
\vspace{-2mm}
\caption{Graph of the kink solution $u(x,t) = 4\tan^{-1}e^{\gamma (x-vt)}$ at time $t=0$ and $t=1$ for the case where $v = 1/2$.}
\label{fig:SGsol1}
\end{center} \end{figure} \vspace{-2mm}
\noindent The choice of the parameter $v$ governs the shape of the graph as $\gamma$ is dependent on $v$ where $v \in (-1, 1)$. At time $t=0$, if $v \rightarrow 1$ or $v \rightarrow -1$ the gradient of the kink increases and the $u$ intercept remains at $\pi$. When $t=1$, $\gamma$ still controls the gradient of the kink, but the positive parameter $v$ also causes the kink to propagate to the right. In Figure 2, $v=1/2$ and so the kink has shifted to the right. If negative $v$ was chosen, then the direction of propagation would be reversed, and the kink would shift to the left.

\subsubsection{$\bm \phi^{4}$ Kinks}
\label{s:s:phi} 
There are other models that can be used to discover other kink solutions. Some examples of these are the $\bm \phi^{4}$ model, Korteweg-de Vries model and the nonlinear Schrödinger model \citep{yousefi2012}. \smallskip \\
If instead the potential function considered is $V(u) = 1/2(1-u^{2})^{2}$ in the Klein-Gordon equation \eqref{kgequ}, then this is known as the {\bf $\bm \phi^{4}$ equation}
\begin{equation}
u_{tt}-u_{xx} - 2u(1-u^{2}) = 0
\label{phi4equ}
\end{equation}
A solution of this equation is the function
\begin{equation}
u(x,t) = \tanh\gamma(x-vt)
\label{phi4sol}
\end{equation}
where $v \in (-1, 1)$ is a constant and $\gamma = (1-v^{2})^{-1/2}$.

\begin{proof} 
In this case the second order partial derivatives can be found as,
\begin{equation*}
u_{xx} = -2\gamma^{2}\tanh\gamma(x-vt)\sech^{2}\gamma(x-vt), \hspace{8mm} u_{tt} = -2\gamma^{2}v^{2}\tanh\gamma(x-vt)\sech^{2}\gamma(x-vt)
\end{equation*}
and an expression for $V'(u) = -2u(1-u^{2})$ can be found as $$-2u(1-u^{2}) = -2\tanh\gamma(x-vt)\sech^{2}\gamma(x-vt)$$
Again after substituting these expressions into \eqref{phi4equ}, we find that $0 = 0$, and hence \eqref{phi4sol} is a solution of the $\phi^{4}$ equation.  
\end{proof} 
\noindent In this equation, the vacua occur when $V(u_{*}) = - 2u(1-u^{2}) = 0$ and solving this equation we obtain the roots $u = \pm1$. This means that the equation \eqref{phi4equ} has vacua occurring at $u_{0}=-1$ and $u_{1}=1$. 
In fact, the solution is actually a {\bf kink} solution as it exhibits the boundary behaviour \eqref{kinkbound}
\begin{equation}
u_{0} = \lim_{x \rightarrow {-\infty}}\tanh\gamma(x-vt) = -1 \hspace{5mm}u_{1} = \lim_{x \rightarrow {\infty}}\tanh\gamma(x-vt) = 1
\label{phi4kinkb}
\end{equation}
Therefore, is it a kink solution of the $\phi^{4}$ equation.

\begin{figure}[H] \begin{center} 
\begin{tikzpicture}
\begin{axis}[xmin=-2.5, xmax=2.5,
    ymin=-1.5, ymax=1.5, samples=1000,
    axis lines=center,
    axis on top=true,
    domain=-2.5:2.5,
    xlabel=$x$, ylabel =$u(x)$, ytick={-1, 0, 1}, legend pos=outer north east]
\addplot+[no marks, red, thick] {tanh(2*sqrt(3)*x/3)};
\addlegendentry{$t=0$}
\addplot+[no marks, teal,thick] {tanh(2*sqrt(3)*(x-1/2)/3)};
\addlegendentry{$t=1$}
\addplot+[no marks, dotted, thick]{-1};  \node [right] at (axis cs: -2.5,-1.1) {$u_{0}$};
\addplot+[no marks, dotted, thick]{1}; \node [right] at (axis cs: -2.5,1.1) {$u_{1}$};
\end{axis}
\end{tikzpicture} \vspace{-1mm} 
\caption{Graph of the kink solution $u(x,t) = \tanh\gamma(x-vt)$ at time $t=0$ and $t=1$ for the case where $v = 1/2$. The choice of the parameter $v$ has the same effect as previously explained for the kink solution of the Sine-Gordon equation.}\label{fig:phi4sol1}
\end{center} \end{figure}
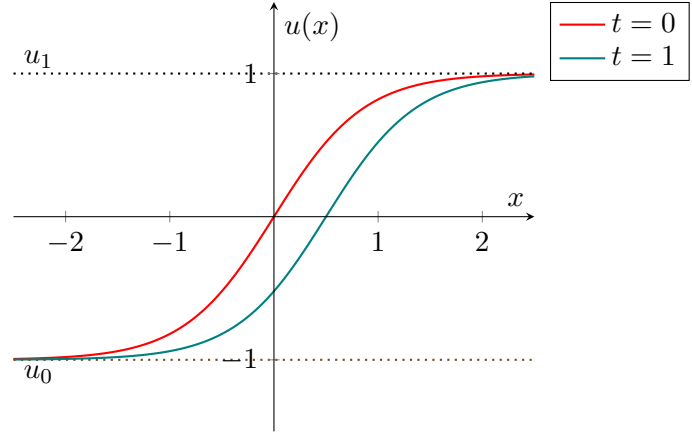  \vspace{-3mm}

\subsection{Conservation Laws}
\label{s:consvlaw}
Given any field $u:$ \field, the {\bf energy} and {\bf momentum} of $u$ are given by
\begin{align}
E(t) &= \infint\left(\frac{1}{2}u_{t}^{2}+\frac{1}{2}u_{x}^{2}+V(u)\right)dx \label{energy}\\ 
P(t) &= -\infint u_{t}u_{x} \;dx \label{mom}
\end{align}
which are both functions of time \citep{fitzgerald2019}.\\
\newline Assume that $u(x,t)$ is a solution of \eqref{kgequ} with finite energy and momentum, then it can be shown that $E(t)$ and $P(t)$ are both {\bf constant} functions:\\
\newline Differentiating $E(t)$ with respect to $t$, using the chain rule, we find
\begin{equation*}
 \frac{dE(t)}{dt} = \infint \left(u_{t}u_{tt}+u_{x}u_{xt}+V'(u)u_{t}\right)dx =\infint \left(u_{t}(u_{tt}+V'(u))+u_{x}u_{xt}\right)dx
\end{equation*}
rearranging equation \eqref{kgequ} gives $u_{xx} = u_{tt}+V'(u)$ which can be substituted above to give
\begin{align*}
 \frac{dE(t)}{dt} &=\infint \left(u_{t}u_{xx}+u_{x}u_{xt}\right) dx \\
&=\infint \left(\pdev{}{x}(u_{t}u_{x})\right) dx \quad \text{(by the product rule)}\\
&=[u_{t}u_{x}]^{\infty}_{-\infty}\\
&= 0
\end{align*}
This is equal to zero because the function $u(x,t)$ is at a vacuum. When $x \rightarrow \pm \infty,\: u(x,t) \rightarrow u_{*}$ where $u_{*}\in\mathbb{R}$. Therefore $u_{x}(x,t)$ and $u_{t}(x,t)$ equal zero, making the derivative of $E(t)$ with respect to $t$ equal zero. Hence, this means that the function $E(t)$ is constant.\\
\newline Similarly, differentiating $P(t)$ with respect to $t$ using the product rule, we find
\begin{equation*}
\frac{dP(t)}{dt} = - \infint \left(u_{tt}u_{x}+u_{xt}u_{t}\right)dx
\end{equation*}
rearranging equation \eqref{kgequ} gives $u_{tt} = u_{xx}- V'(u)$ which can be substituted in to give
\begin{align*}
\frac{dP(t)}{dt} &= - \infint \left(\:(u_{xx}-V'(u)\:)u_{x}+u_{xt}u_{t}\right) dx\\
&= - \infint \left(u_{x}u_{xx} - u_{x}V'(u) +u_{xt}u_{t}\right) dx\\
&= - \infint \pdev{}{x}\left[\frac{1}{2}u_{x}^{2}+\frac{1}{2}u_{t}^{2}-V(u)\right] dx \quad \text{(by the chain rule)}\\
&= -\left[\frac{1}{2}u_{x}+\frac{1}{2}u_{t}-V(u)\right]^{\infty}_{-\infty}\\
&= 0 
\end{align*}
This equals zero by the same argument used for $E(t)$ and recalling that $V(u_{*}) = 0$ for a vacuum. Hence, the function $P(t)$ is also a constant function. \\
\newline The integrand in the expression defining $E$,
\begin{equation}
\mathscr{E}(x,t) = \frac{1}{2}u_{t}^{2}+\frac{1}{2}u_{x}^{2}+V(u)
\label{engden}
\end{equation}
is the {\bf energy density} of the field $u$ \citep{fitzgerald2019}.

\subsubsection{Sine-Gordon Potential}
\label{s:s:sgpot}
The energy of the SG kink \eqref{sGsol} can be computed by finding $u_{x}, u_{t}, V(u)$ and substituting these into \eqref{energy},
$$u_{t} =\frac{-4\gamma ve^{\gamma(x-vt)}}{e^{2\gamma(x-vt)}+1} \qquad u_{x} = \frac{4\gamma e^{\gamma(x-vt)}}{e^{2\gamma(x-vt)}+1} \qquad V(u) = 1 - \cos\left(4\tan^{-1}e^{\gamma(x-vt)}\right)$$
We can split the energy formula \eqref{energy} into three parts to make the calculation easier \citep{rem1999}.
\begin{equation}
E_{sg} = E_{1} + E_{2} + E_{3} 
\label{energy123}
\end{equation}
{\bf Note:} We can use the following identity to simplify the calculations,
\begin{equation}
\sech^{2}(x) = 4/(e^{2x}+e^{-2x} + 2)
\label{sechsqiden}
\end{equation}
For the kinetic energy term $E_{1}$,
\begin{align*}
E_{1} = \infint \frac{1}{2}u_{t}^{2}\;dx  &= \frac{1}{2} \infint \frac{4\gamma^{2}v^{2}e^{2\gamma(x-vt)}}{e^{2\gamma(x-vt)}}\left(\frac{4}{e^{2\gamma(x-vt)} + e^{-2\gamma(x-vt)}+2}\right)\;dx \\
&= \frac{1}{2} \infint 4\gamma^{2}v^{2}\sech^{2}\gamma(x-vt)\;dx\\
&=2\gamma v^{2} \infint \sech^{2}(x)\;dx
\end{align*}
where we have accounted for the $\gamma$ term inside the integral by taking out a factor of $1/\gamma$, and absorbing the $vt$ term since integrating with limits $\pm\infty$ means that the shift it represents is already accounted for. So,
\begin{equation}
E_{1} = 2\gamma v^{2} [\tanh(x)]^{\infty}_{-\infty} = 2\gamma v^{2} [1 - (-1)] = 4\gamma v^{2}
\label{eng1}
\end{equation}
For $E_{2}$, we can do a similar calculation using the same identity \eqref{sechsqiden} to obtain,
\begin{equation}
E_{2} = \infint \frac{1}{2} u_{x}^{2}\;dx  = \frac{1}{2}\infint 4\gamma^{2}\sech^{2}\gamma(x-vt)\;dx =2\gamma \infint \sech^{2}(x) \;dx= 4\gamma
\label{eng2}
\end{equation}
For the last term $E_{3}$, we can use the same approach as in \ref{s:s:sgkinks} and use \eqref{definezy} and \eqref{sincosy} from the right angled triangle to get $\cos(y) = 1/\sqrt{z^{2}+1}$. Then using the double angle formula, $\cos(4y) = 1 - 2\sin^{2}(2y)$ we find,
\begin{equation*}
E_{3} = \infint V(u) \;dx = \infint 2\sin^{2}(2y) \;dx
\end{equation*}
We can use the half-angle formula $\sin(2x) = 2\tan(x)/\left(1+\tan^{2}(x)\right)$ and then substitute the expression we used for $y$ to write this as,
\begin{equation*}
E_{3} = 2\infint \left(\frac{2\tan(y)}{1+\tan^{2}(y)}\right)^{2} \;dx = 2\infint \left(\frac{2e^{\gamma(x-vt)}}{1+e^{2\gamma (x-vt)}}\right)^{2}\;dx
\end{equation*}
We can then simplify this using the identity $\sech(x) = 2/e^{x}+e^{-x}$ to arrive at,
\begin{equation}
E_{3} = 2\infint \sech^{2}\gamma(x-vt) \;dx= \frac{2}{\gamma} \infint \sech^{2}(x) \;dx = \frac{4}{\gamma}
\label{eng3}
\end{equation}
Finally, substituting all three terms \eqref{eng1}, \eqref{eng2} and \eqref{eng3} into \eqref{energy123}, the total energy of a SG kink is 
\begin{equation}
E_{sg} = 4\gamma v^{2} + 4\gamma +\frac{4}{\gamma} = 8\gamma
\label{energysg}
\end{equation} 
We can also calculate the momentum of the SG kink by substituting $u_{x}$ \& $u_{t}$ into \eqref{mom} and again use \eqref{sechsqiden},
\begin{equation}
P_{sg} = \infint \frac{-16\gamma^{2}ve^{2\gamma(x-vt)}}{e^{4\gamma(x-vt)}+2e^{2\gamma(x-vt)}+1} \;dx = -4\gamma v \infint \sech^{2}(x) \;dx = -4\gamma v \times 2 =  -8\gamma v
\label{momsg}
\end{equation}
The equations \eqref{energysg} \& \eqref{momsg} are exactly the relativistic energy and momentum equations that appear in special relativity \citep{wood2003}.
Here the rest mass of the soliton is $m_{0} =8$ and $v$ is the velocity relative to an observer. Relativistic momentum and energy are conserved in the same way that classical momentum and energy are. \smallskip \\
It can also be found, using \eqref{gamma}, that the number $E_{sg}^{2} - P_{sg}^{2}$ is {\em independent} of the parameter $v$,
\begin{equation}
E_{sg}^{2} - P_{sg}^{2} = (8\gamma)^{2} - (8\gamma v)^{2} = 64\gamma^{2}(1-v^{2}) = 64
\label{engmomrelsg}
\end{equation}
In special relativity, the relativistic energy-momentum relation is
\begin{equation}
E^{2} - |P|^{2}c^{2} = m_{0}^{2}c^{2}
\label{specialrel}
\end{equation}
where $c$ is the speed of light and $m_{0}$ is the rest mass \citep{wood2003}. \smallskip \\
Usually \eqref{gamma} is defined to be $\gamma = (1- (v/c)^{2})^{-1/2}$, where $v$ is the speed of a moving observer and $c$ is the speed of light, but we have chosen to set $c=1$ for simplicity. So, \eqref{specialrel} becomes 
\begin{equation}
E^{2} - P^{2} = m_{0}^{2}
\label{engmomrel}
\end{equation}
Therefore the equation \eqref{engmomrelsg} satisfies this relation where the rest mass is $m_{0} = 8$. If the momentum was zero, the equation \eqref{specialrel} is just the extended version of Einstein's famous $E =mc^{2}$ that has been adapted for the case where the energy is relative but the mass is still the rest mass, rather than the relative mass. This is a very surprising and interesting result to obtain from the soliton wave solution that we began with. It supports the fact that the solitons really do have particle-like features, such as a rest mass, and further reinforces the idea of wave-particle duality. Therefore, we can interpret kinks and antikinks as smoothed out relativistic particles and antiparticles respectively \citep{fitzgerald2019}. 

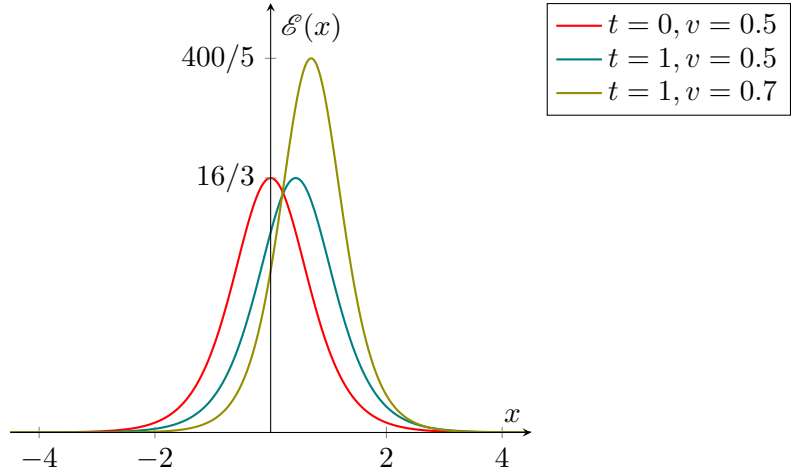
\begin{figure}[H] \begin{center}
\begin{tikzpicture}
\begin{axis}[xmin=-4.5, xmax=4.5,
    ymin=0, ymax=9, samples=1000,
    axis lines=center,
    axis on top=true,
    domain=-4.5:4.5,
    xlabel=$x$, ylabel=$\mathscr{E}(x)$, ytick={16/3, 400/51}, yticklabels={$16/3$, $400/5$}, legend pos=outer north east]
\addplot+[no marks, red, thick] {16/3*1/(cosh(2*sqrt(3)*x/3))^2};
\addlegendentry{$t=0, v=0.5$}
\addplot+[no marks, teal,thick] {16/3*1/(cosh(2*sqrt(3)*x/3-1/2))^2};
\addlegendentry{$t=1, v=0.5$}
\addplot+[no marks, olive,thick] {(400/51)*1/(cosh(1.400280084*(x-0.7))^2};
\addlegendentry {$t=1, v=0.7$}
\end{axis}
\end{tikzpicture}  \vspace{-1mm} 
\caption{Energy density $\mathscr{E}(x)$ =  $4\gamma^{2}\sech^{2}\gamma(x-vt)$ of the SG kink solution at times $t=0$ and $t=1$ for the case where $v=0.5$ and $v=0.7$ } 
\label{fig:sgengd} \end{center} \end{figure}  \vspace{-4mm} 
\noindent The energy density of the kink is strongly localised and peaked about its centre. Since $v \in (-1,1)$, as the velocity approaches $1$ or $-1$, the width of the energy density profile decreases and contracts. An example can be seen above in figure \ref{fig:sgengd} represented by the green line.

\subsubsection{$\bm \phi^{4}$ Potential}
The energy of the $\phi^{4}$ kink solution \eqref{phi4sol} can be computed by finding $u_{x}, u_{t}, V(u)$ and substituting these into \eqref{energy} as follows,
$$u_{t}=-\gamma v \sech^{2}\gamma(x-vt) \qquad u_{x}=\gamma \sech^{2}\gamma(x-vt) \qquad V(u) = \frac{1}{2} \sech^{4}\gamma(x-vt)$$
\begin{equation*}
E_{\phi^{4}}= \frac{1}{2}(\gamma^{2}(v^{2}+1)+1)\infint \sech^{4}\gamma(x-vt) \;dx =\gamma\infint \sech^{4}(x) \;dx
\end{equation*}
where we have used the same simplification techniques that were used for the SG potential. This leads to,
\begin{equation}
E_{\phi^{4}} = \gamma\left[\tanh(x)\right]_{-\infty}^{\infty} - \gamma\left[\frac{\tanh^{3}(x)}{3}\right]_{-\infty}^{\infty}=\gamma\left[2- \frac{2}{3}\right]=\frac{4}{3}\gamma
\label{engphi}
\end{equation} 
which is a constant, as expected.\smallskip\\
Similarly, the momentum of the $\phi^{4}$ kink solution \eqref{phi4sol} can be computed by substituting $u_{t}$ and $u_{x}$ into \eqref{mom} and applying the same techniques,
\begin{equation}
P_{\phi^{4}} = \gamma^{2}v \infint \sech^{4}\gamma(x-vt)\; dx =\gamma v \infint \sech^{4}(x)\; dx =\frac{4}{3}\gamma v
\label{momphi}
\end{equation}
It can also be found that the number $E_{\phi^{4}}^{2} - P_{\phi^{4}}^{2}$ is {\em independent} of the parameter $v$ 
\begin{equation}
E_{\phi^{4}}^{2} - P_{\phi^{4}}^{2} = \left(\frac{4}{3}\gamma\right)^{2} -  \left(\frac{4}{3}\gamma v\right)^{2} = \frac{16}{9}\gamma^{2}(1-v^{2}) = \frac{16}{9}
\label{engmomrelphi}
\end{equation}
using \eqref{gamma}.  
From \eqref{engmomrel}, the rest mass of the $\phi^{4}$ kink solution is $m_{0} = 4/3$.

\begin{figure}[H] \begin{center}
\begin{tikzpicture}
\begin{axis}[xmin=-2.5, xmax=2.5,
    ymin=0, ymax=1.5, samples=1000,
    axis lines=center,
    axis on top=true,
    domain=-2:3,
    xlabel=$x$, ylabel=$\mathscr{E}(x)$, ytick={4/3}, yticklabels={$4/3$}, legend pos= north east]
\addplot+[no marks, red, thick] {4/3*1/(cosh(2*sqrt(3)*x/3))^4};
\addlegendentry{$t=0$}
\addplot+[no marks, teal,thick] {4/3*1/(cosh(2*sqrt(3)*x/3-1/2))^4};
\addlegendentry{$t=1$}
\end{axis}
\end{tikzpicture}
\caption{Localised Energy density, $\mathscr{E}(x)$ =  $\gamma^{2}\sech^{4}\gamma(x-vt)$ of the $\phi^{4}$ kink solution, peaked about its centre at times $t=0$ and $t=1$ for the case where $v=1/2$.} 
\end{center} \end{figure}
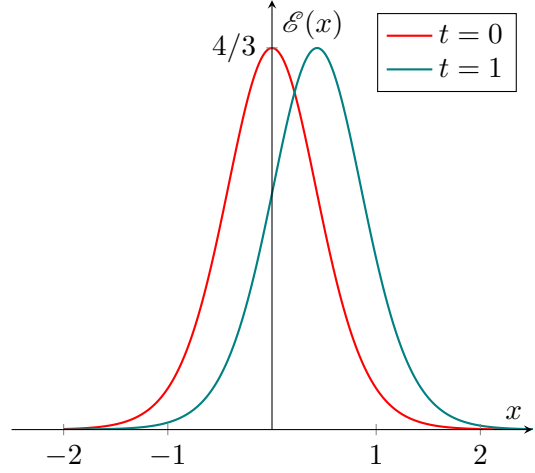

\subsection{Static Kinks and Antikinks}
Static kinks are kink solutions of \eqref{kgequ} that do not depend on time, $t$.  This means, $u(x)$ is a static kink if and only if it has kink boundary behaviour \eqref{kinkbound} and satisfies the {\em ordinary} differential equation (ODE)
\begin{equation}
\frac{d^{2}u}{dx^{2}} = V'(u) 
\label{static}
\end{equation}
This can be reduced to a {\em first order} differential equation by writing the equation as a derivative of $x$ and integrating
\begin{align*}
\left(\frac{du}{dx}\right)\frac{d^{2}u}{dx^{2}} &= \frac{dV}{du}\left(\frac{du}{dx}\right)\\
\frac{d}{dx}\left(\frac{1}{2}\left(\frac{du}{dx}\right)^{2}\right) &= \frac{dV(u(x))}{dx}\\
\frac{1}{2}\left(\frac{du}{dx}\right)^{2} &= V(u(x))
\end{align*}
which can be rearranged to obtain the first order differential equation
\begin{equation} 
\frac{du}{dx} = \pm \sqrt{2V(u)}
\label{static1st}
\end{equation}
This equation represents the gradient of $u(x)$. Hence, $$\frac{du}{dx} = +\sqrt{2V(u)} \qquad \text{and} \qquad \frac{du}{dx} = -\sqrt{2V(u)}$$ respectively represent a {\bf static kink}, where the gradient is positive and a {\bf static antikink}, where the gradient is negative. As the kink is static, the energy equation \eqref{energy} is now $E= \infint \frac{1}{2}u_{x}^{2}+V(u)\;dx$, therefore \eqref{static1st} means that the energy is divided equally between the elastic energy represented by the first term and the potential energy represented by the second term. \smallskip\\
As static kinks/antikinks are time independent, the energy density is given by
$$\mathscr{E}(x,t) = \frac{1}{2}u_{x}^{2}+V(u)$$
The first order ODE \eqref{static1st} can be substituted into this to obtain
\begin{align*} 
\mathscr{E}(x) &= \frac{1}{2}\left(\pm\sqrt{2V(u)}\right)^{2} + V(u)\: =\: 2V(u) \:=\: \left(\frac{du}{dx}\right)^{2}\; \text{(using \eqref{static1st} rearranged)}
\end{align*}
Hence, the energy density of a static kink/antikink is given by
\begin{equation}
\mathscr{E}(x) = u_{x}^{2}
\label{engdstatic}
\end{equation}

\subsubsection{Sine-Gordon Potential}
If the SG model \eqref{sGequ} is taken into consideration, we can plot the potential function $V(u) = 1 - cos(u)$ 
\begin{figure}[H] \begin{center}
\begin{tikzpicture}
\begin{axis}[xmin=-6.5, xmax=6.5,
    ymin=-0.5, ymax=2.5, samples=1000,
    axis lines=center,
    axis on top=true,
    domain=-6.5:6.5,
    xlabel=$x$, ylabel =$u(x)$, xtick={-6.28, -3.14, 0, 3.14, 6.28}, ytick={2},  xticklabels ={$-2\pi$, $-\pi$, 0, $\pi$, $2\pi$}, yticklabels = {$2$}, legend pos=outer north east]
\addplot [mark=none,draw=red, thick]{1-cos(deg(x)))};
\end{axis}\end{tikzpicture}
\caption{Potential function $V(u) = 1- \cos(u)$ of the SG model with vacua at $u = 2\pi n$, where $n \in \mathbb{Z}$}
\end{center} \end{figure}
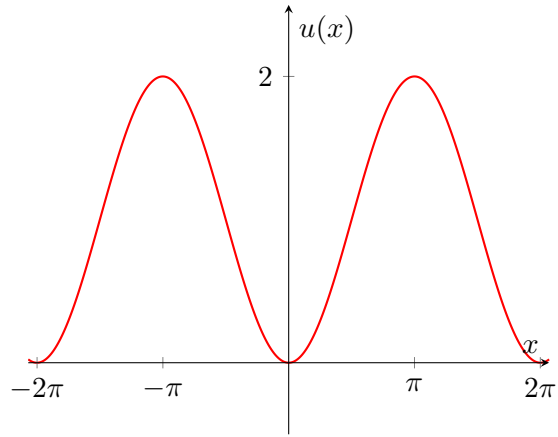
\noindent The static kink solution of \eqref{kgequ} with this potential can be found by using \eqref{static1st},
\begin{equation*}
\frac{du}{dx} = \pm \sqrt{2} \sqrt{1-\cos(u)}
\end{equation*}
\begin{equation}
\int \frac{1}{\sqrt{1-\cos(u)}}\;du = \pm \int \sqrt{2} \;dx
\end{equation}
Then we can substitute the square root of the half angle formula $2\sin^{2}(x/2) = 1 - \cos(x)$, and rearrange to obtain,
\begin{equation}
\int \frac{1}{\sin\left(\frac{u}{2}\right)}\;du = \pm \int 2\;dx
\end{equation}
Integrating and simplifying this leads to,
\begin{equation}
\lnn{\tan\frac{u}{4}} =\pm x + c_{1}
\end{equation}
where $c_{1}$ is a constant. \smallskip \\
This can then be rearranged to obtain the static SG kink solution $u(x) = 4\tan^{-1}e^{\pm x + c_{1}}$ with boundary behaviour \eqref{sgkinklim}.
The energy density of this solution, using \eqref{engdstatic} \& \eqref{sechsqiden} is, 
\begin{equation}
\mathscr{E}(x) = u_{x}^{2} = \left(\frac{4e^{x+c_{1}}}{e^{2(x+c_{1})}+1}\right)^{2} = \frac{16e^{2(x+c_{1})}}{(e^{2(x+c_{1})}+1)^{2}} = 4\sech^{2}(x+c_{1})
\end{equation}
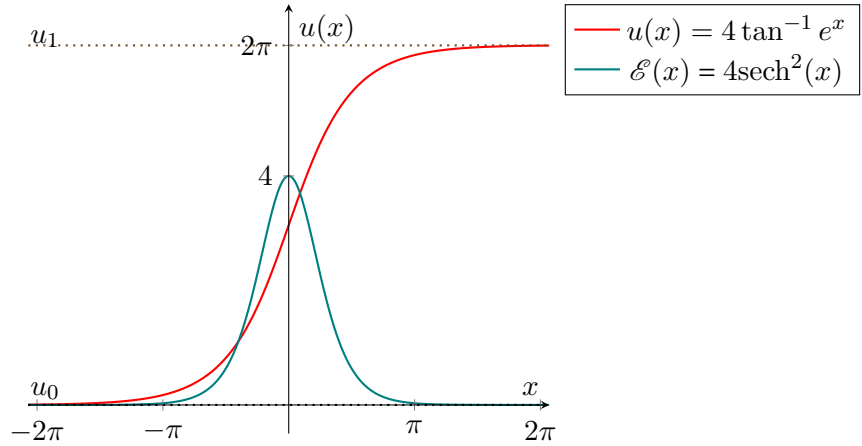
\begin{figure}[H] \begin{center}
\begin{tikzpicture}
\begin{axis}[xmin=-6.5, xmax=6.5,
    ymin=-0.5, ymax=7, samples=1000,
    axis lines=center,
    axis on top=true,
    domain=-6.5:6.5,
    xlabel=$x$, ylabel =$u(x)$, xtick={-6.28, -3.14, 0, 3.14, 6.28}, ytick={4, 6.28},  xticklabels ={$-2\pi$, $-\pi$, 0, $\pi$, $2\pi$}, yticklabels = {$4$, $2\pi$}, legend pos=outer north east]
\addplot[mark=none,draw=red, thick]{4*rad(atan(e^(x))};
\addlegendentry{$u(x) = 4\tan^{-1}e^{x}$}
\addplot [mark=none,draw=teal, thick]{4*(1/cosh(x))^2};
\addlegendentry{$\mathscr{E}(x)$ = $4\sech^{2}(x)$}
\addplot+[no marks, dotted, thick]{6.28};  \node [right] at (axis cs: -6.7,6.4) {$u_{1}$};
\addplot+[no marks, dotted, thick]{0}; \node [right] at (axis cs: -6.7,0.25) {$u_{0}$};
\end{axis}\end{tikzpicture}
\caption{Static SG kink solution and its energy density $\mathscr{E}(x)$ where $c_{1} =0$}
\end{center} \end{figure}

\subsubsection{$\phi^{4}$ Potential}
If instead the $\phi^{4}$ model is now considered, we can plot the potential function $V(u) = \frac{1}{2}(1-u^{2})^{2}$ 

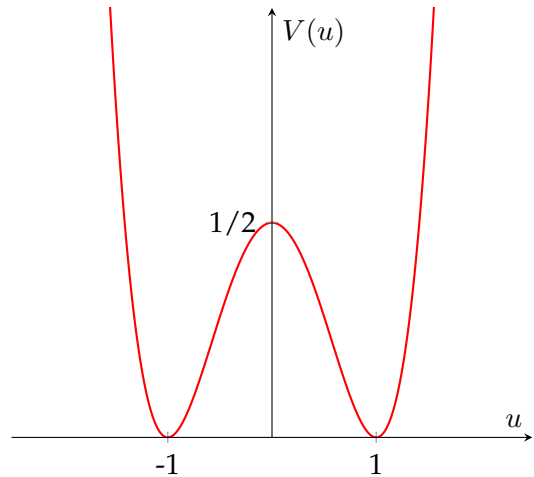
\begin{figure}[H] \begin{center}
\begin{tikzpicture}
\begin{axis}[ xmin=-2.5, xmax=2.5, ymin=0, ymax=1, samples=1000, axis lines=center, axis on top=true, domain=-2.5:2.5, ylabel=$V(u)$, xlabel=$u$, xtick={-1,0,1}, xticklabels ={-1, 0, 1}, ytick={1/2}, yticklabels ={1/2}]
\addplot [mark=none,draw=red, thick] {1/2*(1-x^2)^2};
\end{axis} \end{tikzpicture}
\caption{Potential function $V(u)= \frac{1}{2}(1-u^{2})^{2}$ of the $\phi^{4}$ model with vacua at $u = -1$ and $u=1$} 
\end{center} \end{figure}
\noindent The static kink and antikink solutions of \eqref{kgequ} with this potential can be found by using \eqref{static1st}.
\begin{equation*}
\frac{du}{dx} = \pm \sqrt{2\left(\frac{1}{2}(1-u^{2})\right)^{2}} = \pm\: 1-u^{2}
\end{equation*}
$$\int \frac{1}{1-u^{2}}\: du = \pm \int dx $$
This first order separable ODE can be solved using the LHS known integral $\int \frac{1}{1-u^{2}} = \tanh(u)$ and so, $u(x) = \tanh(x \pm c_{1}) $, using the positive sign in the RHS and where $c_{1}$ is an arbitrary constant. This is a static {\bf kink} solution as it satisfies the $\phi^{4}$ kink boundary behaviour \eqref{phi4kinkb}.\\
Similarly, if the negative sign in the RHS of the equation is used, then we obtain the solution $u(x) = \tanh(-x \pm c_{2})$, where $c_{2}$ is an arbitrary constant. This is a static {\bf antikink} solution as the $\phi^{4}$ kink boundary behaviour is now reversed. \\

The energy density of the static kink solution, using \eqref{engdstatic}, is calculated as
\begin{equation}
\mathscr{E}(x) = u_{x}^{2} = \left(\sech^{2}(x+c_{1})\right)^{2} = \sech^{4}(x+c_{1})
\end{equation}

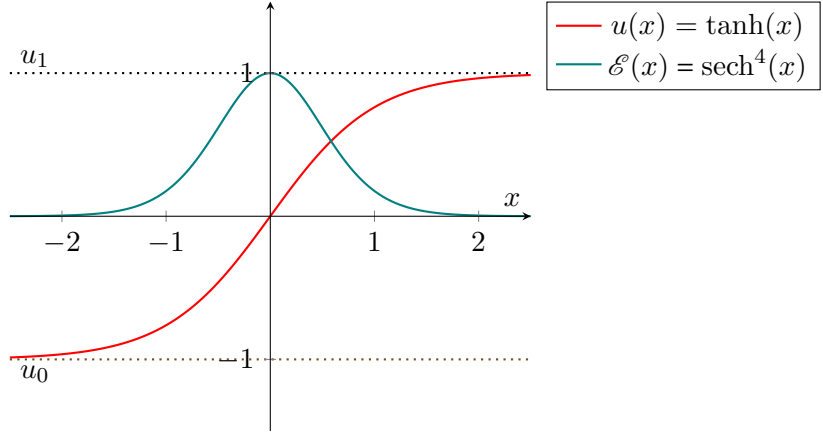
\begin{figure}[H] \begin{center}
\begin{tikzpicture}
\begin{axis}[xmin=-2.5, xmax=2.5, ymin=-1.5, ymax=1.5, samples=1000, axis lines=center, axis on top=true, domain=-2.5:2.5, xlabel=$x$, legend pos=outer north east]
\addplot+[no marks, red, thick] {tanh(x)};
\addlegendentry{$u(x)=\tanh(x)$}
\addplot+[no marks, teal,thick] {1/(cosh(x))^4};
\addlegendentry{$\mathscr{E}(x)$ = $\sech^{4}(x)$}
\addplot+[no marks, dotted, thick]{-1};  \node [right] at (axis cs: -2.5,-1.1) {$u_{0}$};
\addplot+[no marks, dotted, thick]{1}; \node [right] at (axis cs: -2.5,1.1) {$u_{1}$};
\end{axis}
\end{tikzpicture}
\caption{Static $\phi^{4}$ kink solution and its energy density $\mathscr{E}(x)$ where $c_{1} = 0$}
\end{center} \end{figure}

\subsection{The Bogomol'nyi Argument}
\label{s:s:bogo} 
If $u_{0} < u_{1}$ are neighbouring vacua of the potential $V$, then $V: [u_{0}, u_{1}] \rightarrow \mathbb{R}$ is positive on $(u_{0}, u_{1})$ and smooth, so there exists an increasing function $F: [u_{0}, u_{1}] \rightarrow \mathbb{R}$ such that 
\begin{equation}
V(u) = \frac{1}{2}F'(u)^{2}
\end{equation} 
with $F'(u_{0}) = F'(u_{1}) = 0$ \citep{fitzgerald2019}.\smallskip \\
The function 
\begin{equation}
F(u) = \intu \sqrt{2V(u)}\; du
\end{equation}
could be used. Note that $F(u_{1}) > F(u_{0})$ as $F$ is increasing. \smallskip \\
If $u(x,t)$ is any field with kink boundary behaviour, then the energy can be calculated as
\begin{align*}
E &= \frac{1}{2}\infint (u_{t}^{2}+u_{x}^{2}+F'(u)^{2}) dx\\
&\geq \frac{1}{2}\infint (u_{x}^{2}+F'(u)^{2}) dx\\
&= \frac{1}{2}\infint (u_{x}-F'(u))^{2} dx + \infint u_{x}F'(u) dx \\
&\text{(by difference of two squares \& adding a correction term)}\\
&\geq \infint u_{x}F'(u) dx\\
&= [F(u(x))]_{-\infty}^{\infty} \quad \text{(after integrating)}\\
&= F(u_{1}) - F(u_{0}) \quad \text{(using kink boundary behaviour)}
\end{align*}
Hence, the energy bound of every field with kink boundary behaviour \eqref{kinkbound} is the {\bf Bogomol'nyi bound}
\begin{equation}
E \geq \; M =  F(u_{1}) - F(u_{0}) = \intu \sqrt{2V(u)}\; du
\label{bogbound}
\end{equation}
If $E=M$ then the non negative terms which were discarded in the chain of inequalities above become zero. In the first inequality, $u_{t} = 0$ since we chose $E=M$, this means that the field is {\em static} as there is no longer any dependence on time. In the second inequality, the term $(u_{x}+F'(u))^{2}$ is also equal to zero using \eqref{static1st}, since the field is static.\medskip\\
Therefore the lower bound, where the total energy is {\em minimised} is the {\bf Bogomol'nyi equation}
\begin{equation}
E = \; M = \intu \sqrt{2V(u)}\; du
\label{bogequ}
\end{equation}
where the field is static and satisfies the static kink equation \eqref{static1st}.\medskip \\
The Bogomol'nyi argument can be adapted to find that fields with antikink boundaries \eqref{akinkbound} obey the same Bogomol'nyi bound \eqref{bogbound} with equality if and only if the field is static and it obeys the static antikink equation.\medskip\\
For the {\bf sine-Gordon} potential $V(u) = 1 - \cos(u)$, introduced in section \ref{s:s:sG}, the energy of every static kink and antikink is 
\begin{equation} 
E = \int_{0}^{2\pi} \sqrt{2(1-\cos(u))}\; du =\; 2\int_{0}^{2\pi} \sin\left(\frac{u}{2}\right)\; du =\; 4\left[-\cos\left(\frac{u}{2}\right)\right]_{0}^{2\pi}=\; 8
\label{staticbogSG}
\end{equation}
For the {\bf phi-4} potential $V(u) = \frac{1}{2}(1-u^{2})^{2}$, introduced in \ref{s:s:phi}, the energy of every static kink and antikink is
\begin{equation}
E = \int_{-1}^{1} (1 - u^{2})\; du =\; \left[u - \frac{u^{3}}{3}\right]_{-1}^{1} = \frac{4}{3}
\label{staticbogphi}
\end{equation}
These are equivalent to \eqref{energysg} \& \eqref{engphi}, where the only difference is that \eqref{gamma} becomes $\gamma = 1$ since the kinks are static so $v=0$.

\subsection{Lorentz Invariance}
\label{s:s: lorentz}
If $u$: \field\: is a field, and any constant $v \in (-1,1)$ is chosen, then a {\em new} field can be defined as $\bar{u}$: \field\: from $u$ by the formula
\begin{equation}
\bar{u}(x,t) = u(\gamma(x-vt), \gamma(t-vx)),
\label{lortrans}
\end{equation}
where $\gamma = (1-v^{2})^{-1/2}$ \citep{fitzgerald2019} is the Lorentz factor that determines the change of relativistic mass, time and length of the kink while it is moving.
The formula \eqref{lortrans} is essentially a transform on $u$ and it can be shown that $u$ satisfies \eqref{kgequ} if and only if $\bar{u}$ does.
\begin{proof}
Define $\bar{x} = \gamma(x-vt) \; \text{and} \; \bar{t} = \gamma(t-vx)$, and so \: $\bar{u}(x,t) = u(\bar{x},\bar{t})$\\
Then the first order partial derivative of $u$ with respect to $x$ using the chain rule and the defined $\bar{x}$ can be found as
\begin{equation*}
\pdev{u}{x} =\; \pdev{u}{\bar{x}}\pdev{\bar{x}}{x} + \pdev{u}{\bar{t}}\pdev{\bar{t}}{x} =\; \gamma\pdev{u}{\bar{x}} - \gamma v \pdev{u}{\bar{t}}
\end{equation*}
This can then be used to find the second order partial derivative with respect to $x$, considering $u$ is just an arbitrary field
\begin{equation*}
\pdev{^{2}u}{x^{2}} = \pdev{}{x}\left(\pdev{u}{x}\right) = \left(\gamma\pdev{}{\bar{x}} - \gamma v \pdev{}{\bar{t}}\right)\left(\gamma\pdev{u}{\bar{x}} - \gamma v \pdev{u}{\bar{t}}\right) = \gamma^{2}\pdev{^{2}u}{\bar{x}^{2}} - 2\gamma^{2}v\frac{\partial {u}}{\partial {\bar{x} \partial{\bar{t}}}} + \gamma^{2}v^{2} \pdev{^{2}u}{\bar{t}^{2}}
\end{equation*}
and similarly, the second order partial derivative of $u$ with respect to $t$ can be found as
\begin{equation*}
\pdev{^{2}u}{t^{2}} = \gamma^{2}v^{2}\pdev{^{2}u}{\bar{x}^{2}} - 2\gamma^{2}v\frac{\partial {u}}{\partial {\bar{x} \partial{\bar{t}}}} + \gamma^{2} \pdev{^{2}u}{\bar{t}^{2}}
\end{equation*}
Substituting the above second order partial derivatives into \eqref{kgequ} and using \eqref{gamma}, the equation
\begin{equation}
u_{\bar{t}\bar{t}} - u_{\bar{x}\bar{x}} + V'(u) = 0
\end{equation}
is obtained. This is exactly the nonlinear Klein-Gordon equation in terms of the new field \eqref{lortrans}. Hence, $u$ satsfies \eqref{kgequ} if and only if $\bar{u}$ does, and therefore the Klein-Gordon equation is {\em invariant} under the Lorentz transform. 
\end{proof}
\noindent If the field $u$ is {\em static} (a function of $x$ only) then
\begin{equation}
\bar{u}(x, t) = u(\gamma(x-vt))
\end{equation} 
is called the {\bf Lorentz boost} of $u$ with velocity $v$ . This boosted kink travels with constant velocity $v$  and it has a profile {\em sharper} than that of the static kink since $\gamma > 1$ if $v \neq 0$ \citep{fitzgerald2019}. This means a travelling kink is narrower than a static one and this phenomenon is called the {\bf Lorentz contraction}. \medskip\\ 
If $u$ is a static kink of energy $M$ and $\bar{u}$  is the Lorentz boost with velocity $v$, the energy $E$ of the boosted static kink is found by defining $\bar{x} = \gamma(x-vt)$ again, so $\bar{u}(x, t) = u(\bar{x}, 0)$. \smallskip\\
Using \eqref{energy}, the energy of a static kink is 
\begin{equation} 
E(t) = M = \infint\left(\frac{1}{2}u_{x}^{2}+V(u)\right)dx 
\label{engstatic}
\end{equation}
The relation $\pdev{u}{x} = \pdev{u}{\bar{x}}\pdev{\bar{x}}{x}=\gamma\pdev{u}{\bar{x}}$ can be substituted into \eqref{engstatic} to obtain
\begin{equation}
E = \gamma\infint\left(\frac{1}{2}u_{\bar{x}}^{2}+V(u)\right)d\bar{x} \qquad \text{(using $d\bar{x}/{dx}= \gamma$)}
\end{equation}
The integral in this equation is equivalent to the energy of a static kink $M$, as the Klein-Gordon equation is invariant under the Lorentz transform. Therefore, the energy of the boosted static kink $\bar{u}$ is 
\begin{equation}
E = \gamma M
\label{boostengstatic}
\end{equation}
The momentum $P$ of the boosted static kink $\bar{u}$ can be found by a similar method, this time using the momentum equation \eqref{mom} and the energy density equation \eqref{engden} to obtain,
\begin{equation}
P = \gamma v M
\label{boostmomstatic}
\end{equation}
It can also be found that
\begin{equation} 
E^{2} - P^{2} = M^{2}  \qquad \text{(for all $v$)}
\end{equation}
as substituting \eqref{boostengstatic} and \eqref{boostmomstatic} as well as using \eqref{gamma} gives
\begin{equation*}
E^{2} - P^{2} = \gamma^{2}M^{2} (1 - v^{2}) = \gamma^{2} M^{2} \left(\frac{1}{\gamma^2}\right) =M^{2}  
\end{equation*}
This equation holds for {\em all} $v$ as it is independent of the parameter $v$ and this $\gamma$ dependent energy and momentum is again precisely analogous to special relativity discussed in section \ref{s:s:sgpot}. We can see that all these relations hold for the static energy of the SG and $\phi^{4}$ kinks, \eqref{staticbogSG} \& \eqref{staticbogphi} respectively. \medskip \\

\noindent If we assume that $v$ is small, then the binomial expansion formula can be used to calculate the energy of the boosted static kink \eqref{boostengstatic}. \\
Using \eqref{gamma}, and expanding for quadratic terms only,
\begin{equation*}
\gamma= (1-v)^{-1/2}(1+v)^{-1/2} =(1+\frac{v}{2}+\frac{3}{8}v^{2}+...)(1-\frac{v}{2}+\frac{3}{8}v^{2}-...) =1+\frac{1}{2}v^{2}+...
\end{equation*} 
where higher order terms $v^{4}$ can be neglected because of the low velocity. Hence, 
\begin{equation} 
E = \gamma M \approx M + \frac{1}{2}Mv^{2}
\label{engboostexp}
\end{equation}
This shows that the boosted static kink has energy equal to the sum of the static kink energy $M$ and the kinetic energy caused by the boost, represented by the second term in \eqref{engboostexp}. As we saw for the SG and $\phi^{4}$ kinks where $v \neq 0$ in section \ref{s:consvlaw}, the energy and momentum for a boosted or travelling {\em static} kink is almost exactly the same, provided the velocity $v$ is small, since the Klein-Gordon equation is Lorentz invariant.

\section{A Closer Look at the Sine-Gordon Model}
\subsection{Multi-Kink Solutions}
\label{s:s: multi}
The Sine-Gordon equation \eqref{sGequ} has even more soliton solutions alongside the kink solution \eqref{sGsol} discussed previously. There are numerous methods that can be used to obtain these solutions. For example, the method of inverse scattering \citep[p.~262]{rem1999} or B\"acklund transforms \citep[p.~118]{manton2004}. Another method is that \eqref{sGequ} can be solved by separating variables $x$ and $t$. Then, the signs of parameters can be chosen in order to give various solutions which behave in ways interesting to study. The choice of parameters leads to multi-kink solutions in the form of kink-kink collisions, kink-antikink collisions and breathers. \smallskip \\

\noindent In general, soliton solutions of the SG equation are in the form 
\begin{equation}
u(x,t) = 4\tan^{-1} \left[\frac{F(x)}{G(t)}\right]
\label{sgForm}
\end{equation}

\noindent This can be seen from the kink solution \eqref{sGsol} where in this case $F(x)=e^{\gamma x}$ and $G(x)= e^{\gamma v t}$. \smallskip \\

\noindent Following closely the method presented by Remoissenet \citep{rem1999}, we can seek solutions in the from above. If we rearrange \eqref{sgForm} we can see that
\begin{equation}
\frac{u}{4} = \tan^{-1} \left[\frac{F(x)}{G(t)}\right] \hspace{5mm} \text{leading to} \hspace{5mm}
\tan\left(\frac{u}{4}\right) = \frac{F(x)}{G(t)} = \phi
\label{phi}
\end{equation}
Using this, the SG equation \eqref{sGequ} can be transformed into
\begin{equation}
\phi_{tt}-\phi_{xx} - \phi\frac{(1-\phi^{2})}{(1+\phi^{2})^{2}} = 0
\label{transformedsg}
\end{equation}
by the use of the identity
\begin{equation*}
\sin(u) = 4\phi\frac{(1-\tan^{2}\phi)}{(1+\tan^{2}\phi)^{2}}=4\phi\frac{(1-\phi^{2})}{(1+\phi^{2})^{2}}
\end{equation*}

\noindent Then we can use \eqref{phi} to find,
\begin{equation*}
\phi_{x} = \frac{F_{x}}{G} \hspace{5mm}
\phi_{xx} = \frac{F_{xx}}{G} \hspace{5mm}
\phi_{t} = -\frac{FG_{t}}{G^{2}} \hspace{5mm}
\phi_{tt} = \frac{2FG_t}{G^{3}} - \frac{FG_{tt}}{G^{2}}
\end{equation*}
and substitute these derivatives into \eqref{transformedsg} to find that
\begin{equation}
(G^{2} + F^{2})\left(\frac{F_{xx}}{F} +\frac{G_{tt}}{G}\right) - 2\left((F_{x})^{2} + (G_{t})^{2}\right) + F^{2} - G^{2} = 0
\label{tsg1}
\end{equation}

\noindent We can then differentiate \eqref{tsg1} with respect to $x$ and again with respect to $t$ using the product rule to obtain,
\begin{equation}
(G^{2})_{t}\left(\frac{F_{xx}}{F}\right)_{x} + (F^{2})_{x}\left(\frac{G_{tt}}{G}\right)_{t} = 0
\label{tsg2}
\end{equation}

\noindent Now we can rearrange and separate the variables $x$ and $t$ in \eqref{tsg2} and equate these to a separation constant $\lambda$. This will give two ordinary differential equations (ODE) that can then be solved. 
\begin{equation*}
\frac{1}{(F^{2})_{x}}\left(\frac{F_{xx}}{F}\right)_{x} =\hspace{1mm} - \frac{1}{(G^{2})_{t}}\left(\frac{G_{tt}}{G}\right)_{t} \hspace{1mm}= \hspace{1.5mm} \lambda
\label{tsg3}
\end{equation*}
where the two ODEs are 
\begin{align}
\left(\frac{F_{xx}}{F}\right)_{x}  &= \lambda(F^{2})_{x} \label{ode1}\\
\left(\frac{G_{tt}}{G}\right)_{t} &= -\lambda(G^{2})_{t} \label{ode2}
\end{align}

\noindent Both of these ODEs can be integrated with respect to $x$ and $t$. They can then be integrated again after multiplying \eqref{ode1} by $2F_{x}$ and \eqref{ode2} by $2G_{t}$ to obtain
\begin{align}
(F_{x})^{2} &= \frac{\lambda}{2}F^{4} + a_{1}F^{2} + a_{2} \nonumber \\
(G_{t})^{2} &= -\frac{\lambda}{2}G^{4} + b_{1}G^{2} + b_{2} \label{tsg5}
\end{align}
where $a_{1}, a_{2}, b_{1}, b_{2}$ are constants of integration. 

\noindent After substitution of equations \eqref{tsg5} into \eqref{tsg1} the relationship between the constants is found as $a_{1} - b_{1} = 1$ \& $a_{2} - b_{2} = 0$. If we let $\lambda/2 = -q^{2}, a_{1} = p^{2}, a_{2} = n^{2}$ then equations \eqref{tsg5} can be written as
\begin{align}
(F_{x})^{2} &= -q^{2}F^{4} + p^{2}F^{2} + n^{2} \nonumber \\
(G_{t})^{2} &= q^{2}G^{4} + (p^{2}-1)G^{2} - n^{2} \label{ode22}
\end{align}

\noindent Finally from these equations various soliton solutions can be found depending on the parameters chosen. A summary of these is given below\citep{rem1999}:
\begin{center}
 \begin{tabular}{|| c | c | c | c ||} 
 \hline
 \multicolumn{3}{|c|}{\bf Parameters} & {\bf Type} \\  [0.5ex] 
 \hline\hline
 $q=0$ & $p > 0$ & $n=0$ & kink \\ 
 \hline
 $q=0$ & $p^{2} > 1$ & $n \neq 0$ & kink-kink collision\\
 \hline
$q \neq 0$ & $p > 1$ & $n=0$ & kink-antikink collision\\
 \hline
 $q \neq 0$ & $p^{2} < 1$ & $n=0$ & breather \\
 \hline
\end{tabular}
\end{center}

\subsubsection{Kink-Kink Interaction}
\label{s:s:kk}
\noindent If the case $q =0, p^{2} > 1, n \neq 0 $ is considered, then \eqref{ode22} become 
\begin{equation*}
(F_{x})^{2} = p^{2}F^{2} + n^{2} \hspace{5mm} (G_{t})_{2} = (p^{2}-1)G^{2} - n^{2}
\end{equation*}
Solving this, by assuming $n=p$ and letting $v = (p^{2} -1)^{1/2}/p$ gives,
\begin{equation*} 
F(x) = \sinh\left(\frac{1}{(1-v)^{1/2}}x\right), \hspace{5mm} G(t) = \frac{1}{v}\cosh\left(\frac{v}{(1-v)^{1/2}}t\right)
\end{equation*}
Then by using \eqref{gamma} and \eqref{sgForm} the {\bf kink-kink soliton} solution is obtained,
\begin{equation}
u(x,t) = 4\tan^{-1}\left[\frac{v\sinh(\gamma x)}{\cosh(\gamma vt)}\right]
\label{kksol}
\end{equation} \citep{draz1983,rem1999}.  \smallskip \\
This represents a kink-kink collision as we can approximately separate the solution into two SG kink solutions, $k_{1}$ \& $k_{2}$, travelling at opposite velocities in the following way,
\begin{equation*}
u(x,t) = 4\tan^{-1}\left[\frac{v\sinh(\gamma x)}{\cosh(\gamma vt)}\right]  \approx 4\tan^{-1}\left(e^{\gamma(x-vt)}\right) -\hspace{1mm} 4\tan^{-1}\left(e^{-\gamma(x+vt)}\right) = k_{1} -k_{2}
\end{equation*}
It must be noted that this breakdown of \eqref{kksol} is not an actual solution of the SG equation, but instead is an approximation when the velocity $v >0$. It can be seen that $k_{1}$ has boundary behaviour \eqref{sgkinklim} and that $k_{2}$ has boundary behaviour, $u_{0} = -2\pi$ \& $u_{1} = 0$. The equation \eqref{kksol} has no static multi-soliton solutions because the two kinks either attract or repel each other. We can see this in the equation \eqref{kksol} because when $v \rightarrow 0$, the equation becomes degenerate. In the case of a kink-kink interaction, the kinks are in fact repelled from one another without any change to their structure. \smallskip \\
To understand the interaction it is helpful to consider the {\bf topological charge} of the kinks and Bogomol'nyi bounds of section \ref{s:s:bogo}. \smallskip \\
The topological charge is 
\begin{equation}
N = \frac{1}{2\pi} \infint u'\;dx
\label{topcharge}
\end{equation}
where $N$ is the difference between the total number of kinks and the total number of antikinks, also called the net number of solitons \citep{manton2004}. \smallskip \\
Using this, the Bogomol'nyi bound for the SG potential \eqref{bogbound} becomes
\begin{equation}
E \geq M = 8 |N|
\label{engtopc}
\end{equation}

\noindent Recall from \eqref{staticbogSG}, a static SG kink has $E = 8$, where we can see from \eqref{engtopc} that $N=1$. However, we have just noted there are no static multi-soliton solutions possible and so $E > 8 |N|$. Therefore in our case, the energy of a two kink soliton, $N=2$, must then satisfy the strict Bogomol'nyi bound $E > 16$. But, the total energy of \eqref{kksol} is the sum of the energy of two kinks, $8 + 8 = 16$ and so the energy will approach $E = 16$ (in the case they are infinitely separated). When the two kinks travel towards each other, the potential energy increases and $E > 16$ as required. However, as energy is conserved and $E=16$ for separated kinks, this extra energy is accounted for by a repulsive force between them causing them to bounce back. \smallskip \\
This force can be calculated by using the static solution as done in \citep{raj1977} or by finding the rate of change of momentum as in \citep[p.~114]{manton2004} to obtain $F = -32e^{-R}$, giving the {\bf interaction energy} between the two kinks as,
\begin{equation}
E_{int} = \int F \;dR =\; 32e^{-R}
\end{equation} 
where $R$ is the separation between both of the kinks \citep{manton2004}. 

\begin{figure}[H] \begin{center}
 \begin{tikzpicture}
\begin{axis}[xmin=-10.5, xmax=10.5,
    ymin=-7, ymax=7, samples=1000,
    axis lines=center,
    axis on top=true,
    domain=-10.5:10.5,
    xlabel=$x$, ylabel =$u(x)$, xtick={-6.28, -3.14, 0, 3.14, 6.28}, ytick={-6.28, -3.14, 3.14, 6.28},  xticklabels ={$-2\pi$, $-\pi$, 0, $\pi$, $2\pi$}, yticklabels = {$-2\pi$, $-\pi$, $\pi$, $2\pi$}, legend pos=outer north east]
\addplot+[no marks, teal,thick] {4*rad(atan(exp(2*sqrt(3)*(x+5)/3))) - (4*rad(atan(exp(2*sqrt(3)*-(x-5)/3))))};
\addlegendentry{$t=-10$}
\addplot[-latex, thick, forget plot]coordinates{(-6,-4) (-4.5,-4)};
\addplot[-latex, thick, forget plot]coordinates{(6,4) (4.5,4)};
\addplot+[no marks, olive, thick] {4*rad(atan(exp(2*sqrt(3)*x/3))) - (4*rad(atan(exp(2*sqrt(3)*-x/3)))) };
\addlegendentry{$t=0$}
\addplot+[no marks, red,thick] {4*rad(atan(exp(2*sqrt(3)*(x-5/2)/3))) - (4*rad(atan(exp(2*sqrt(3)*-(x+5/2)/3))))};
\addlegendentry{$t=5$}
\addplot[-latex, thick, forget plot]coordinates{(-1.6,-2) (-3.1,-2)};
\addplot[-latex, thick, forget plot]coordinates{(1.6,2) (3.1,2)};
\end{axis}
\end{tikzpicture}  \vspace{-3.5mm}
\caption{Two kinks propagating in opposite directions with velocity $v=\pm1/2$ when $t=-10$ and colliding at the origin when $t=0$. Following the collision, both kinks repel and travel in the opposite direction at $v=\mp 1/2$ showed at $t=5$, and eventually back to their initial starting positions.} 
\label{fig:kkcol} \end{center}  \end{figure}
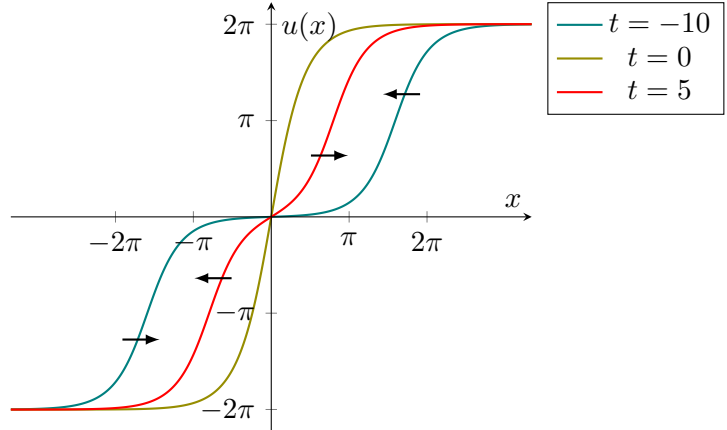  \vspace{-3mm}
\noindent The fact that the interaction energy is {\em positive} indicates that the force between the colliding kinks is repulsive and so the kinks scatter backwards. We can see that $u(x,t) = u(x, -t)$ and so the solution \eqref{kksol} is and even and symmetric about $t=0$, as shown in \ref{fig:kkcol}. Furthermore, from \ref{s:s:sgpot} the interaction between two kinks can be regarded as two relativistic particles colliding and repelling each other due to their identical topological charges \eqref{topcharge}. The two particles could both be positively charged, or both be negatively charged to cause this type of repulsion. Moreover, the kinks are unchanged and survive after the collision, which is exactly how two particles in a collision would behave.

There is another interpretation that can be considered which is that forward scattering occurs instead of backward scattering \citep{sutcliffe1993}. In this conclusion, the two kinks are said to pass through each other rather than repel, by the notion that they accelerate and undergo a shift. From the graph, this interpretation is also valid. The fact that there are two possible interpretations which are both correct also has applications in quantum mechanics, since the kinks can be regarded as indistinguishable particles. Both interpretations are reasonable, but from the interaction energy and the link between solitons behaving like particles, it is more plausible that kink-kink collisions are repulsive.

\subsubsection{Kink-Antikink Interaction}
\label{s:s:ka}
The interaction of a kink and an antikink can be considered by solving the exercise (6.8) presented by Drazin \citep[p.~104]{draz1983}. \\
An exact solution of the SG equation \eqref{sGequ} is 
\begin{equation}
u(x,t) = 4\tan^{-1}\left[\frac{v\cosh(\gamma x)}{\sinh(\gamma vt)}\right]
\label{kasol}
\end{equation}

\begin{proof}
If we use the general form of SG soliton solutions \eqref{sgForm}, we can identify that
\begin{equation}
F(x) = v\cosh(\gamma x) \hspace{3mm} \& \hspace{3mm} G(t) = \sinh(\gamma vt)
\label{kasplit}
\end{equation}
The second order partial derivatives for \eqref{kasplit} with respect to $x$ and $t$ can be found as,
\begin{equation*}
F_{xx} = \gamma^{2}v \cosh(\gamma x) \hspace{5mm} G_{tt} = \gamma^{2}v^{2} \sinh(\gamma vt)
\end{equation*}
Substituting these expressions into \eqref{tsg2},
\begin{equation*}
\frac{d}{dt}(\sinh^{2}(\gamma vt))\frac{d}{dx}(\gamma^{2}) + \frac{d}{dx}(v^{2}\cosh^{2}(\gamma x))\frac{d}{dx}(\gamma^{2}v^{2}) = 0
\end{equation*}
Therefore, \eqref{kasol} is an exact {\bf kink-antikink soliton} solution of the SG equation. 
\end{proof}

\noindent Then we can examine the asymptotic behaviour as $t \rightarrow \pm \infty$.
The solution represents a kink-antikink collision as we can again approximately separate the solution, but this time into a SG kink and an antikink, $k_{3}$ \& $k_{4}$, travelling at opposite velocities in the following way,
\begin{equation}
u(x,t) = 4\tan^{-1}\left[\frac{v\cosh(\gamma x)}{\sinh(\gamma vt)}\right] \approx 4\tan^{-1}\left(-e^{\gamma(x+vt)}\right) -\hspace{1mm} 4\tan^{-1}\left(e^{-\gamma(x-vt)}\right) = k_{3} - k_{4}
\end{equation}
where $k_{3}$ is an anti-kink with boundary behaviour $u_{0} = 0, u_{1} = -2\pi$ and $k_{4}$ is a kink with boundary behaviour $u_{0} = -2\pi, u_{1} = 0$.\smallskip \\
In the same way as the kink-kink solution, the kink-antikink solution \eqref{kasol} is also degenerate when velocity $v=0$ and so no static solutions are possible. However, in this case the, the potential energy decreases as the kink and antikink attract each other. For this interaction, the net number of solitons is $N=0$ as there is one kink and one antikink. Therefore, from \eqref{engtopc}, when the kink and antikink are infinitely separated, $E \geq 0$. Since there are no static multisoliton solutions, $E \neq 0$, hence the total energy is $E > 0$. As the total energy is conserved, this extra energy is now accounted for by an attractive force. \smallskip \\ 
Similar to the kink-kink force, the kink-antikink force is instead {\em positive}, $F = 32e^{-R}$, giving the interaction energy between the kink and antikink as,
\begin{equation}
E_{int} = \int F \;dR =\; -32e^{-R}
\end{equation}

\begin{figure}[H] \begin{center}
\begin{tikzpicture}
\begin{axis}[xmin=-10.5, xmax=10.5,
    ymin=-7, ymax=7, samples=1000,
    axis lines=center,
    axis on top=true,
    domain=-10.5:10.5,
    xlabel=$x$, ylabel =$u(x)$, xtick={-6.28, -3.14, 0, 3.14, 6.28}, ytick={-3.14, -6.28, 3.14, 6.28},  xticklabels ={$-2\pi$, $-\pi$, 0, $\pi$, $2\pi$}, yticklabels = {$-\pi$, $-2\pi$, $\pi$, $2\pi$}, legend pos=outer north east]
\addplot+[no marks, teal,thick] {4*rad(atan((0.5*cosh(2*sqrt(3)*x/3))*(1/(sinh(sqrt(3)/3*-10)))))};
\addlegendentry{$t=-10$}
\addplot+[no marks, red,thick] {4*rad(atan((0.5*cosh(2*sqrt(3)*x/3))*(1/(sinh(sqrt(3)/3*-5)))))};
\addlegendentry{$t=-5$}
\addplot+[no marks, blue,thick] {2*3.1415};
\addlegendentry{$t=0$}
\addplot+[no marks, olive,thick] {4*rad(atan((0.5*cosh(2*sqrt(3)*x/3))*(1/(sinh(sqrt(3)/3*5)))))};
\addlegendentry{$t=5$}
\addplot+[no marks, orange,thick,solid] {4*rad(atan((0.5*cosh(2*sqrt(3)*x/3))*(1/(sinh(sqrt(3)/3*10)))))};
\addlegendentry{$t=10$}
\addplot[-latex, thick, forget plot]coordinates{(-7,-4.4) (-5.5,-4.4)};
\addplot[-latex, thick, forget plot]coordinates{(7,-4.4) (5.5,-4.4)};
\addplot[-latex, thick, forget plot]coordinates{(-3.5,-3) (-2,-3)};
\addplot[-latex, thick, forget plot]coordinates{(3.5,-3) (2,-3)};
\addplot[-latex, thick, forget plot]coordinates{(-5.5,4.4) (-7,4.4)};
\addplot[-latex, thick, forget plot]coordinates{(5.5,4.4) (7,4.4)};
\addplot[-latex, thick, forget plot]coordinates{(-2,3) (-3.5,3)};
\addplot[-latex, thick, forget plot]coordinates{(2,3) (3.5,3)};
\end{axis}\end{tikzpicture}
\caption{Kink and an anti-kink travelling towards each other at opposite velocities $v=\pm 1/2$, annihilating at $t=0$ and then reappearing, with boundary behaviour increased by $+2\pi$ respectively and still travelling at the original velocity.}
\label{fig:kacol} \end{center}\end{figure}
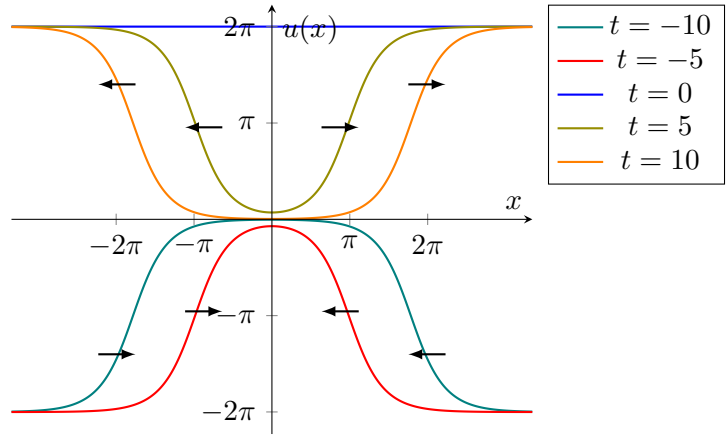  \vspace{-3.5mm}
\noindent The interaction energy being {\em negative} indicates that the force causing the collision is attractive. The exercise (6.8) prompts us to note that the solution ``is an even function of $x$, an odd function of $t$ and is instantaneously a constant at $t=0$'' \citep[p.~104]{draz1983}. This tells us that $u(x,t) = u(-x,t)$, so the solution is symmetric about $x=0$ and $u(x, -t) = -u(x,t)$, meaning the solution is also symmetric about the origin. Both of these features are seen in figure \ref{fig:kacol} and we can also see that at $t=0$, the solution becomes a constant $u = 2\pi$.

An important observation is that ``the kink and antikink scatter elastically'' \citep[p.~123]{manton2004}. This means that the total kinetic energy of both solitons remains the same before and after the interaction. This is an interesting aspect of the collision as $k_{3}$ \& $k_{4}$ do not completely annihilate each other as one would expect from the particle nature of solitons. In particle-antiparticle collisions, the two attract each other due to opposite topological charges \eqref{topcharge}, completely annihilate and produce photons (discrete packets of energy), to conserve the total energy in the system. The figure \ref{fig:kacol} shows that another two soliton solution, of equal energy, reappears again after $t=0$ because of these conservative properties. This is precisely the solution that satisfies the condition that \eqref{kasol} is an odd function of $t$. 
The alternative interpretation is that the kink and the antikink both accelerate from the attraction and shift, increasing their boundary conditions respectively. Like the kink-kink interaction, both these interpretations are valid, however the notion that solitons behave as particles again allows the conclusion that the annihilation explanation is most likely.

\subsection{Continuous Breathers}
\subsubsection{Static Breathers}
A breather soliton is different from the multi-kink collisions we have discussed so far. It is an exact stationary wave solution (ie. when $v=0$), which is spatially localised and also time periodic \citep{malo2014}. The general concept is that this breather solution is the point where in a kink-antikink collision, the two solitons have attracted each other as expected, but neither the kink nor antikink have enough energy to separate themselves back into their original components. The parameter value $p^{2}$ in \eqref{ode22} is important as it enables the distinction between the two interactions \ref{s:s:kk} \& \ref{s:s:ka} from a breather.

For kink-kink and kink-antikink interactions, $p^{2}>1$ \& $p>1$ respectively in \eqref{ode22}. In this case, $p$ is large enough that the two solitons have enough energy to overcome the force holding them together. However for a breather, the parameter $p^{2}$ is bound, as is must be strictly less than $1$. This is why breathers can be described as a ``kink-antikink bound state'' \citep[p.~123]{manton2004} in which the kink and antikink are `stuck' together and are unable to change their shape following the collision. \smallskip \\
To find a breather solution using the separation of variables procedure presented in section \ref{s:s: multi}, we can continue to follow the method by choosing the parameters $q \ne 0, p^{2} <1$ and $n = 0$. With this choice of parameters, equations \eqref{ode22} become 
\begin{equation}
(F_{x})^{2} = -q^{2}F^{4} + p^{2}F^{2} \hspace{8mm} (G_{t})^{2} =q^{2}G^{4} + (p^{2}-1)G^{2} \label {ode44}
\end{equation}
Then defining the following relations, $\alpha^{2} = p^{2}/q^{2}$ \& $\beta = \sqrt{1-p^{2}}/q$,
Equations \eqref{ode44} transform into
\begin{align}
(F_{x})^{2} &= q^{2} (\alpha^{2}F^{2} - F^{4}) \nonumber \hspace{4mm} \text{leading to} \hspace{4mm} 
\frac{1}{F\sqrt{\alpha^{2} - F^{2}}} dF = \pm q\;dx \\
(G_{t})^{2} &= q^{2}(G^{4} - \beta^{2}G^{2}) \hspace{4mm} \text{leading to} \hspace{4mm} \frac{1}{G\sqrt{G^{2}-\beta^{2}}} dG = \pm q\;dt \label{odebreather}
\end{align}
By using the following known integrals, we can integrate the ODEs \eqref{odebreather} to obtain $F(x)$ and $G(t)$.  \medskip \\
In general for $R = A + Bx + Cx^{2}$,
\begin{align*}
\int \frac{1}{x\sqrt{R}}\; dx &= - \frac{1}{\sqrt{A}}\lnn{\frac{\sqrt{R} + \sqrt{A}}{x}+\frac{B}{2\sqrt{A}}} \hspace{3mm} \text{for $A > 0$} \\
\int \frac{1}{x\sqrt{R}}\; dx &= \frac{1}{\sqrt{-A}}\sin^{-1}\left(\frac{Bx+2A}{x\sqrt{B^{2}-4AC}}\right) \hspace{5mm} \text{for $A<0$}
\end{align*}
So for our case, integrating \eqref{odebreather} using these given integrals where $B=0$ gives,
\begin{equation}
-\frac{1}{\alpha} \lnn{\frac{\sqrt{\alpha^{2} - F^{2}} +\alpha}{F}} = \pm qx + c_{1} \hspace{8mm} \frac{1}{\beta}\sin^{-1}\left(\frac{\beta}{G}\right) = \pm qt + c_{2}
\label{odebrethint}
\end{equation}
where $c_{1}$ and $c_{2}$ are constants of integration which can both be set to equal zero. \smallskip \\
These can now be solved for $F(x)$ and $G(t)$ to obtain,
\begin{equation}
F(x) = \alpha \sech(\alpha qx) \hspace{8mm} G(t) = \pm \frac{\beta}{\sin(\beta qt)} \label{fgbreather}
\end{equation} 
where in the first equation we have used the trigonometric identity $\sech^{-1}(x) = \lnn{1+\sqrt{1-x^{2}}/x}$.\\
\noindent Now we can substitute $\alpha, \beta$ \& \eqref{fgbreather} into \eqref{sgForm} to get,
\begin{equation*}
u(x,t) = 4\tan^{-1} \left[\frac{F(x)}{G(t)}\right] = 4\tan^{-1}\left[\frac{p}{\sqrt{1-p^{2}}} \frac{\sin(\sqrt{1-p^{2}} t)}{\cosh(pt)}\right]
\end{equation*}
Then finally letting $\omega  = \sqrt{1-p^{2}}$ we can write this as,
\begin{equation}
u(x,t) = 4\tan^{-1}\left[\frac{\sqrt{1-\omega^{2}}}{\omega} \frac{\sin(\omega t)}{\cosh(\sqrt{1-\omega^{2}} x)}\right]
\label{breathersol}
\end{equation}
which is the {\bf breather soliton} solution with frequency $\omega \in [-1,1]$ \citep{rem1999}. \medskip\\ 

\begin{figure}[H] \begin{center}
\begin{tikzpicture}
\begin{axis}[xmin=-10.5, xmax=10.5,
    ymin=-7, ymax=7, samples=1000,
    axis lines=center,
    axis on top=true,
    domain=-10.5:10.5,
    xlabel=$x$, ylabel =$u(x)$, xtick={-6.28, -3.14, 0, 3.14, 6.28}, ytick={-3.14, -6.28, 3.14, 6.28},  xticklabels ={$-2\pi$, $-\pi$, 0, $\pi$, $2\pi$}, yticklabels = {$-\pi$, $-2\pi$, $\pi$, $2\pi$}, legend pos=outer north east]
\addplot+[no marks, blue,thick]{4*rad(atan((sqrt(1-0.04)/0.2)*(sin(0.2*-100)*(1/cosh(sqrt(1-0.04)*x))))};
\addlegendentry{$t=-100$}
\addplot+[no marks, teal,thick]{4*rad(atan((sqrt(1-0.04)/0.2)*(sin(0.2*50)/cosh(sqrt(1-0.04)*x))))};
\addlegendentry{$t=-50$}
\addplot+[no marks, red,thick]{4*rad(atan((sqrt(1-0.04)/0.2)*(sin(0.2*-50)/cosh(sqrt(1-0.04)*x))))};
\addlegendentry{$t=50$}
\addplot+[no marks, orange,thick]{4*rad(atan((sqrt(1-0.04)/0.2)*(sin(0.2*100)/cosh(sqrt(1-0.04)*x))))};
\addlegendentry{$t=100$}
\end{axis}\end{tikzpicture}
\caption{Breather oscillations at various times where frequency is fixed at $\omega = 0.2$ and $\omega \in [-1,1]$.}
\label{fig:breather} \end{center}\end{figure}
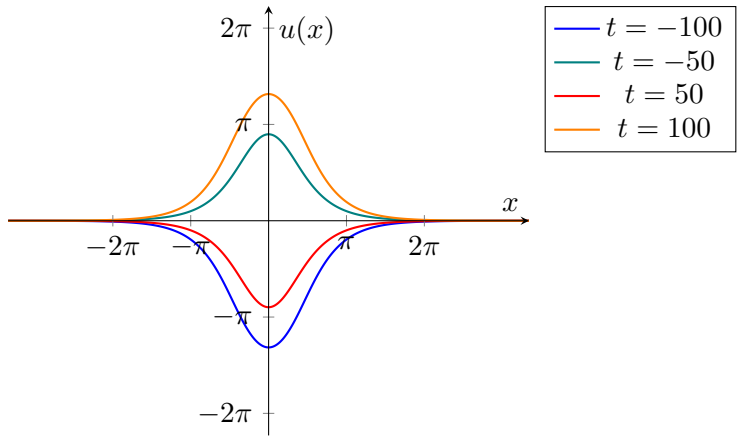  \vspace{-3.5mm}
\noindent As well as oscillating periodically, the amplitude of the wave varies with time when frequency is fixed. We can see from the solution \eqref{breathersol} that the time period is $T = 2\pi/\omega$ and the amplitude is $A = 4\tan^{-1}(\sqrt{1-\omega^{2}}/\omega)$. The solution becomes a constant $u=0$ when $\omega = -1, \omega = 1$ or $t =0$. If time is fixed, dialling the frequency $\omega$ up from $-1$ to $0$ causes the amplitude to increase with every oscillation and so the profile changes from a very tightly bound breather where $p^{2} \ll 1$ to a more loosely bound one where $p^{2}<1$ in \eqref{ode22}. As $\omega$ is increased, the width of the soliton profile also increases and decreases which imitates a breathing motion, which explains why they are called 'breathers'. There is no solution when $\omega =0$ since \eqref{breathersol} would be degenerate. As omega increases from $0$ to $1$, the amplitude gradually decreases with each periodic oscillation.

This solution essentially explains what happens at the kink-antikink collision which occurred at $t=0$ where we saw \eqref{kasol} become instantaneously a constant in figure \ref{fig:kacol}. It was unusual that the two solitons did not completely annihilate each other, and a breather solution explains why this was the case. This is easier to understand by a motion picture of a breather and how the solution behaves on a ribbon \citep{genz1980}. In this video, a kink and an antikink approach each other from opposite sides of the ribbon and when the waves collide, they combine and pause at the peak (maximum positive amplitude). Following this pause, the combined solitons swing down to the trough (maximum negative amplitude) and then have sufficient energy to separate and continue travelling on the ribbon. However as we mentioned before the breather does not have enough energy to separate and so the combined kink and antikink continue to oscillate periodically on the ribbon, stuck in the bound state. 

\subsubsection{Moving Breathers \& Breather Energy}
As the SG equation is invariant under the Lorentz transformation discussed in section \ref{s:s: lorentz}, the breather soliton solution \eqref{breathersol} can also be boosted using \eqref{lortrans} as,

\begin{equation}
u(x,t) = 4\tan^{-1}\left[\frac{\sqrt{1-\omega^{2}}}{\omega} \frac{\sin(\omega \gamma (t-vx)}{\cosh(\sqrt{1-\omega^{2}} \gamma (x-vt)}\right]
\label{boostedb}
\end{equation}
This represents a travelling wave solution with velocity $v<1$. Similar to the boosted kink, the breather profile becomes narrower and sharper due to the Lorentz contraction \citep{rem1999}.  \smallskip \\
We can use \eqref{engtopc} to see that the total energy is $E \geq 0$, since the net number of solitons in this case is $N =0$. Even though the breather case is similar to the kink-antikink interaction \ref{s:s:ka} and static breathers do exist, $E \neq 0$ as the kink and the antikink are not infinitely separated, since they are bound together. Hence $E > 0$ for the moving breather as well as the static breather. So if energy is inserted into the system less than the rest mass of two kinks, we obtain a bound kink-antikink pair, i.e. a breather. We can calculate the energy of the moving \& static breather by using \eqref{boostengstatic} to obtain,
\begin{equation}
E_{B} = 2M\gamma = 16\gamma \sqrt{1-w^{2}} 
\label{breathereng}
\end{equation}
where $M = 8$ is the energy of a static SG kink/antikink. \smallskip \\
We can then see that the binding energy which holds the kink and antikink together in a bound state is given by the difference between the energy of an infinitely separated kink-antikink pair and the energy of a breather  \citep{rem1999}. 
\begin{equation}
E_{bind} = 2M - E_{B} = 16 \gamma (1 - \sqrt{1-w^{2}})
\end{equation} 
This is the {\em maximum energy} that a breather could possibly have. This is because if the system had any more energy and $p^{2} >1$ in \eqref{ode22}, the kink and antikink that the breather is made up of would be able to separate from each other and no longer be a breather solution.
In the case of the Lorentz boosted breather \eqref{boostedb}, we can also use \eqref{boostmomstatic} to find the momentum of the moving breather with velocity $v$,
\begin{equation}
P_{B} = 2M\gamma v = 16\gamma v \sqrt{1-w^{2}}
\label{breathermom}
\end{equation}

\section{Discrete Sine-Gordon Model}
\subsection{Crystal Lattice}
The solitons that we have studied so far can all be visible in the macroscopic world. For example, they can be seen in water waves, like the initial discovery by Russell \citep{Russell1845} or surprisingly can even form in the snow.
\begin{figure}[H] \begin{center}
\includegraphics[scale =0.75]{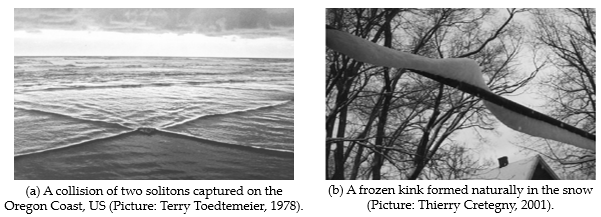}
\caption{Examples of solitons created in nature. Images taken from \citep[p.~24,51]{daux2006}.}
\end{center} \end{figure}
\noindent If instead we study solitons at the microscopic level, we can consider a nonlinear lattice that models these discrete microscopic structures \citep{rem1999}. This one dimensional lattice structure is made up of a one dimensional chain of $n$ particles. \smallskip \\
{\bf Note:} We are using the notion that the solitons behave as particles and so the solution represents the field of atomic displacements $u_n$  \\
\begin{figure}[H] \begin{center}
\includegraphics[scale =0.5]{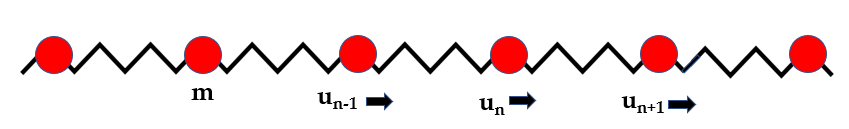}
\caption{1D chain of particles where amplitude is indicated by arrows, recreated using \citep{rem1999}.}
\end{center} \end{figure}
\noindent This is an example of the particles in equilibrium as each are evenly spaced from one another. They interact with their nearest neighbour and all have a mass $m$. They behave in the same way as coupled oscillators, which are masses connected by springs. Longitudinal waves are created when energy is inserted into the system. The longitudinal displacement of the $n^{th}$ particle from its original equilibrium is $u_{n}$, and so the relative displacement between the particles is $u_{n+1} - u_{n}$.

\subsection{Frenkel-Kontorova model \& The Discrete SG Equation}

\noindent The Frenkel-Kontorova model is one that describes a chain of one-dimensional nearest neighbour interacting scalar oscillations. The {\bf Hamiltonian} of this model (ie. the function describing the total energy in the system) is given by,
\begin{equation}
H = \sum\limits_{n} \left[\frac{1}{2}{{P}_{n}}^{2} + V(u_{n}) + W(u_{n+1} - u_{n})\right]
\end{equation}
where integer $n$ is the lattice site number of the chain, $u_{n} = u_{n}(t)$ is the time dependant coordinate of the $n^{th}$ particle and $P_{n}$ is the associated momentum \citep{flach2012, nij2019}. \smallskip \\
The first term represents the {\bf kinetic energy} as $\dot{u} = P_{n}$, the second term represents the {\bf on-site/substrate potential} and the third represents the {\bf interaction potential}. We can substitute the discrete periodic SG potential as the on-site potential and the interaction potential respectively as,
\begin{equation}
V(u_{n}) = 1 - \cos(u_{n}),  \hspace{7mm} W(u_{n+1} - u_{n}) = \frac{1}{2h^{2}}(u_{n+1} - u_{n})^{2}
\end{equation}
to arrive at the Hamiltonian for the one-dimensional SG crystal lattice,
\begin{equation}
H = \sum\limits_{n} \left[\frac{1}{2}{{P}_{n}}^{2} + \frac{1}{2h^{2}}(u_{n+1} - u_{n})^{2} +  (1 - \cos(u_{n}))\right]
\label{sgham}
\end{equation}
where $u_{n}$ is now the displacement of the $n^{th}$ particle from an initial point of coordinate $x =nh$ and $h$ is the discreteness parameter that controls the lattice spacing \citep{dmit2000}. \smallskip \\

The equations of motion \citep{nij2019} from \eqref{sgham} are \vspace{2mm}
\begin{equation}
\vspace{-4.5mm}
\hspace{-7.4cm}
\dot{U}_{n} = \pdev{H}{P_{n}} = P_{n} 
\label{motion1}
\end{equation}

\begin{align}
\dot{P}_{n} =  -\pdev{H}{U_{n}} &= -W'(u_{n+1} - u_{n}) +W'(u_{n} - u_{n-1}) - V'(u_{n}) \nonumber  \\
&= \frac{1}{h^{2}} (u_{n-1} - 2u_{n} +u{n+1}) - \sin(u_{n}) 
\label{motion2}
\end{align}
Combining these gives the {\bf discrete sine-Gordon equation},
\begin{equation}
\ddot{u}_{n} - \frac{1}{h^{2}} (u_{n-1} - 2u_{n} + u_{n+1}) + \sin(u_{n}) = 0
\label{discsg}
\end{equation} 
\citep{dmit2000}. \\
\noindent This discrete SG equation was introduced by Frenkel \& Kontorova in 1938 where they explored crystal dislocations. The equation became even more popular to study because of the continuum limit $h \rightarrow 0$. If $h \rightarrow 0$ in \eqref{discsg} then the Frenkel-Kontorova model becomes the {\em continuous} sine-Gordon equation \eqref{sGequ}, for which we discussed the exact solutions possible. Unlike the continuous SG equation, the discrete equation is not exactly integrable\citep{benner2000} and omits no analytical solutions, therefore it can only be solved numerically. However, it is still interesting to study because breather solitons can still be found. \smallskip \\

\subsection{Linearised Discrete SG Equation}
The equations of motion can be linearised around the ground state $U_{n} = P_{n} = 0$ to find coupled linear equations that can be solved to obtain
\begin{align}
U_{n}(t) &\sim e^{i(\kappa n - \omega_{\kappa}t)} \label{phonon} \\
\omega_{\kappa}^{2} &= 1 + \frac{4}{h^{2}}\sin^{2}\left(\frac{\kappa}{2}\right) \label{dispersion}
\end{align}
where $0 \leq \kappa \leq \pi$ is the wave number and $\omega_{\kappa}$ is the associated frequency \citep{dmit2000}. \smallskip \\
Similar to the continuum case phonons in \ref{s:linsg}, equation \eqref{phonon} represents plane waves of small amplitude, and \eqref{dispersion} is the {\bf dispersion relation} between the wave number $\kappa$ and the frequency $\omega_{\kappa}$ of the phonons. 
As the system is constructed on a crystal lattice, this frequency $\omega_{\kappa}$ \hspace{0.5mm} is periodically dependent on $\kappa$. \smallskip \\
From \eqref{dispersion} the phonon frequency is given by,
\begin{equation}
\omega_{\kappa} = \sqrt{1 + \frac{4}{h^{2}}\sin^{2}\left(\frac{\kappa}{2}\right)}
\label{phofreq}
\end{equation}
\noindent As $ 0 \leq \kappa \leq  \pi$, the phonon spectrum ranges between $\omega_{0} \leq \omega_{\kappa} \leq \omega_{\pi}$, which is called the phonon band. Using \eqref{phofreq} we can calculate the minimum and maximum phonon frequencies respectively as 
\begin{equation}
\omega_{0} = 1 \hspace{5mm} \text{and} \hspace{5mm} \omega_{\pi} = \sqrt{1 + 4/h^{2}}
\end{equation}
The maximum phonon frequency is also known as the Debye cut off frequency \citep{flach2012}. \smallskip \\
Recall from the linearised continuous SG model \ref{s:linsg} we found the dispersion relation as \eqref{clindisprel}. This is a long wave expansion of the discrete SG dispersion relation \citep{benner2000}. As phonons are the {\em linear} excitations with small amplitudes, when the amplitude is large enough, and exceeds the phonon cut off frequency, the linearised approximation \eqref{phonon} is no longer correct \citep{benner2000}. When the frequency $\omega_{\pi} > \sqrt{1 + 4/h^{2}}$, the excitations are now classed as {\em nonlinear} and represent the kinks and breather solutions we have studied in the continuum limit. 

\subsection{Discrete Breathers}
Discrete breathers are also known as Intrinsic Localised Models (ILM) which have internal oscillations. These oscillations are sometimes called envelope oscillations\citep{rem1999} as they contain a fixed amount of energy and can be formed on a 1D crystal lattice.
\begin{figure}[H] \begin{center}
\includegraphics[scale =0.5]{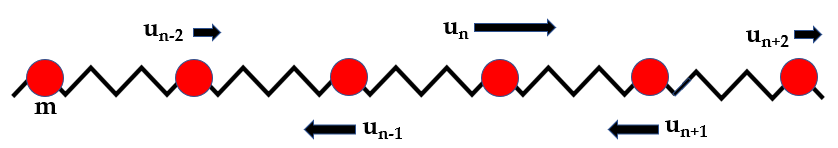}
\caption{Schematic diagram of a discrete breather, recreated using \citep{rem1999}.}
\label{fig:dblattice}
\end{center} \end{figure}
\noindent The figure \ref{fig:dblattice} shows a series of particles which are equally spaced from each other. The particles are oscillating with an amplitude represented by arrows. The length of the arrow indicates the size of the amplitude and the displacement from the equilibrium, so the oscillations decrease spatially from the particle $u_{n}$. The direction of the arrows indicate that each neighbouring particle is vibrating out of phase with each other. 
\begin{figure}[H] \begin{center}
\includegraphics[scale =0.4]{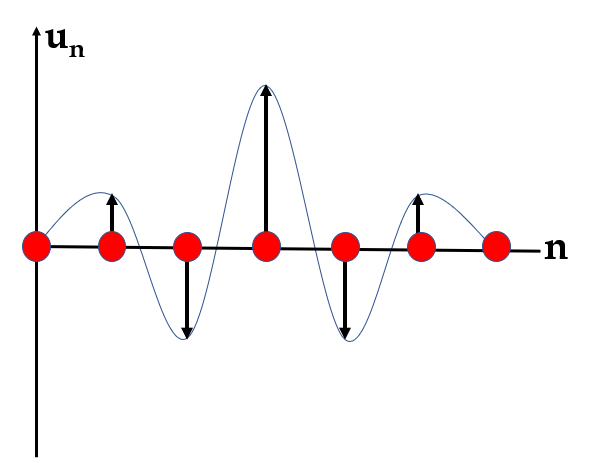}
\caption{Schematic diagram of the longitudinal displacement corresponding to the particles in \ref{fig:dblattice}, recreated using \citep{rem1999}, where the blue line is a visual aid.}
\label{fig:dbdisp}
\end{center} \end{figure}

\noindent Continuous breathers can only exist in a limited number of systems, however discrete breathers are ``structurally stable as soon as nonlinear oscillators are coupled sufficiently weakly and locally'' \citep[p.~251]{rem1999}. The breather solution changes from large amplitude nonlinear waves to low amplitude linear waves, depending on the internal frequency. Therefore, this means that if the linear phonon frequency band is not coupled strongly to the nonlinear breather frequency band, then the breather will be stable and will not dissipate. The term 'coupled' means that the linear and nonlinear waves are transferring energy to one another, so in this case the phonon and breather are not transferring energy.  \smallskip \\

\noindent The lifetime that the breather has can be increased if the breather frequency is greater than a certain value. This value, shown by Boesch and Peyrard \citep{boe1991,dmit2000} is,
\begin{equation}
\omega > \frac{1}{3}\omega_{\pi}
\label{maxlifeb}
\end{equation}
where $\omega$ is the breather frequency. \smallskip \\
However, we know that $\omega \in [-1,1]$ for a breather, and so the frequency cannot be above $1$. Therefore, the breather frequency band is 
\begin{equation}
\frac{1}{3}\omega_{\pi} < \omega < 1
\label{breatherfreqbound}
\end{equation}
This is solved to obtain a value for the discreteness parameter $h$ as
\begin{equation}
h > \frac{\sqrt{2}}{2}
\label{hvalue}
\end{equation}
Usually, if energy was inserted into the lattice, the waves would dissipate which is exactly the case with phonons. If the phonon and breather frequency bands are overlapping, then all energy from the breather would eventually spread amongst the phonons that are emitted.  Whereas, when \eqref{maxlifeb} and \eqref{hvalue}, the breather frequency is uncoupled and so the breather retains its energy. This spatial localisation of the breather is due to two properties. The first is the {\em discreteness} of the system, provided by the underlying lattice and the second is the {\em nonlinearity} of the SG potential function. Because of this, discrete breathers have many applications, some of which are in Josephson Junction networks, ultracold atoms in optical lattices and localised atomic vibrations in molecules and solids \citep{gorb2008}.

\newpage
\section{Summary}
Solitons are wave solutions to nonlinear PDEs, particularly the Klein Gordon equation was interesting to study as different potentials could be used to find different soliton solutions. In the integrable SG model the three elementary types of excitations were explored – Phonons, Kinks and Breathers. From the relativistic properties of this model, evidence of particle-like behaviour of solitons was uncovered. We found that because of the integrability of the system, solitons had energy and momentum that was constant and always conserved. The section on static kinks showed that the SG equation could be simplified into an ODE and an energy bound called the Bogomol’nyi bound can be used to find the energy of every static kink. The fact that the SG model was Lorentz invariant meant that moving kinks and breathers existed and that their energy and momentum for a boosted static kink was the same as the usual kink with $v \neq 0$. 

The various multi-kink solutions to the SG model were found by the use of a separation of variables method. In this method, the type of soliton solution that was obtained, either a kink-kink, kink-antikink or a breather was dependent on a choice of parameters. It was found that kink-kink collisions behave similarly to two identical particles colliding by repelling from each other unchanged. For kink-antikink collisions, the opposite occurred and the solitons were attracted towards each other, like a particle and an antiparticle with opposite charges would. We found that if the kink-antikink pair had enough energy, then they would be able to separate back into their constituent parts, otherwise they would be stuck in a bound state known as a breather. Moreover, the Frenkel Kontorova model that described a chain of oscillating particles on a one-dimensional crystal lattice was the SG model in the continuum limit. The equations of motion of the FK model, with the right choice of potentials led us to the discrete SG model. From the linearization of this discrete model, phonon and breather frequency bands were found and these explained the stability of the breather. If the two frequency bands were not overlapping, then the linear small amplitude phonons and nonlinear large amplitude waves would not couple strongly. This meant that the breather could increase its lifetime and retain its energy without dissipating.

Overall, solitons are highly interesting to study and this report only explored some of the one-dimensional solutions possible. The theory could be extended to vortices and lumps in 2D, monopoles and Skyrmions in 3D or even instantons in 4D.

\newpage


\end{document}